\documentclass[prb,aps,amsmath,amssymb,nofootinbib,epsfig,showpacs,twocolumn,eqsecnum,superscriptaddress]{revtex4-1}

\usepackage[colorlinks,plainpages=false,linkcolor=black,urlcolor=blue,citecolor=blue,pdfpagemode=UseNone,pdfstartview=FitH]{hyperref}

\usepackage{amsmath,cleveref,amssymb,amsfonts,graphicx,multirow,color,bbm,textcomp,mathtools,dsfont}
\usepackage{paralist}
\usepackage{srcltx}
\usepackage[all,cmtip]{xy}
\usepackage{mathrsfs}
\usepackage{srcltx,soul,subfigure}
\usepackage{txfonts}
\usepackage{cleveref}

\newcommand{\bee}{\begin{equation}}
\newcommand{\ee}{\end{equation}}
\newcommand{\bma}{\begin{pmatrix}}
\newcommand{\ema}{\end{pmatrix}}
\newcommand{\balig}{\begin{align}}
\newcommand{\ealig}{\end{align}}

\newcommand{\ba}{\begin{align}}
\newcommand{\ea}{\end{align}}

\newcommand{\ignore}[1]{}

\usepackage{siunitx}

\usepackage{longtable}
\newcommand{\ket}{\rangle}
\newcommand{\bra}{\langle}

\usepackage{dsfont}

\begin{document}
\title{Interaction effects on the classification of crystalline topological insulators and superconductors}

\author{Xue-Yang Song}
\affiliation{International Center for Quantum Materials, School of Physics, Peking University, Beijing, 100871, China}
\affiliation{Max-Planck-Institut f\"ur Festk\"orperforschung,
  Heisenbergstrasse 1, D-70569 Stuttgart, Germany} 
\author{Andreas P. Schnyder}
\affiliation{Max-Planck-Institut f\"ur Festk\"orperforschung,
  Heisenbergstrasse 1, D-70569 Stuttgart, Germany} 

\date{\today}

\begin{abstract}

We classify interacting topological insulators and superconductors with order-two crystal symmetries (reflection or twofold rotation), focusing on the case where interactions reduce the noninteracting classification.
We find that the free-fermion $\mathbb{Z}_2$ classifications are stable against quartic 
contact interactions, whereas the $\mathbb{Z}$ classifications reduce to $\mathbb{Z}_N$,
where $N$ depends on the symmetry class and the dimension $d$.
These results are derived 
using a quantum nonlinear $\sigma$ model (QNLSM) that describes the effects of 
the quartic interactions on the boundary modes of the crystalline topological phases.
We use Clifford algebra extensions to derive the target spaces of these QNLSMs in a unified way.
The  reduction pattern of the free-fermion classification then follows from the presence or absence of topological terms in
the QNLSMs, which is determined by the homotopy group of the target spaces.  
We show that this derivation can be performed using either a complex fermion or a real Majorana representation
of the crystalline topological phases and demonstrate that these two representations give consistent results.
To illustrate the breakdown of the noninteracting classification we present
examples of crystalline topological insulators and superconductors in dimensions one, two, and three, whose
surfaces modes are unstable against  interactions.
For the three-dimensional example, we show that the
reduction pattern obtained by the QNLSM method agrees with the one
inferred from the stability analysis of the boundary modes using bosonization.


%

\end{abstract}
\maketitle

\section {Introduction}

In recent years, the field of topological quantum matter has seen rapid advances~\cite{hasan2010,qi2011,chiu2015,ludwig2016}, stimulated by the discovery of topological insulators~\cite{Bernevig1757,koenig766,hsieh_nature_08} and by potential applications in device fabrication~\cite{pesin_macDonald_TI_applications} 
and quantum information technology~\cite{sarma_nayak_application_QCP}.
An important concept in this field is the notion of symmetry protected topological (SPT) quantum states, which are short-range entangled gapped phases with a symmetry. A defining property of SPT states is that they cannot be deformed to a trivial state by a symmetry-preserving deformation without closing the gap.
 One of the main characteristics of SPT states  is the existence of protected gapless surface states, which leads to many interesting phenomena, such as 
dissipationless currents in two-dimensional systems and magneto-electric effects in three-dimensional topological insulators.
 
An important theme in the field of topological matter is the classification of SPT phases, i.e., to determine how many distinct SPT states exist for a given set of symmetries. For free-fermion systems with nonspatial symmetries (such as, time-reversal) a systematic classification was obtained and summarized in the so-called periodic table of topological insulators (TIs) and superconductors (TSCs)~\cite{schnyder2008,ryu2010,kitaev2009,schnyder_AIP_09}. This table, which is sometimes called the ``ten-fold way'', categorizes $d$-dimensional free-fermion systems into ten symmetry classes~\cite{altlandZirnbauerPRB97,Zirnbauer:1996fk} distinguished by the presence or absence of time-reversal, particle-hole, or chiral symmetry. 
It was shown that in any dimension $d$ there exist five symmetry classes with non-trivial SPT states, that can be indexed  
 by the Abelian groups $\mathbb{Z}$ or $\mathbb{Z}_2$.
 Subsequently, this classification scheme was extended to non-interacting SPT phases with crystalline space group symmetries (i.e., spatial symmetries)~\cite{fu_TCI_PRL11,shiozaki2014,chiu2013,lu2014,morimoto2013,ando_fu_review_15,slager13}, which are important in many condensed-matter systems. There are a number of materials
 which have recently been proposed as candidates for crystalline topological insulators. Among them are the rocksalt SnTe~\cite{Tanaka:2012fk,Hsieh:2012fk,Xu:2012} and the antiperovskites $A_3$PbO~\cite{Hsieh_antiperovskite_12,chiu_antiperovskite_2016}, where~$A$ denotes an alkaline earth metal.

While the classification of free-fermion SPT states is  quite well understood, attention has now shifted to interacting SPT
phases. The motivation to study strongly correlated SPT quantum states comes in part from a number of $5d$- and $4f$-electron systems,
that could be interacting topological insulators.  These include iridium oxide materials~\cite{nagaosa_iridium_oxide},
transitionmetal heterostructures~\cite{xiao_nat_commun_11}, and the Kondo insulator SmB$_6$~\cite{dzero_coleman_review}.
Interactions can modify the classification of free-fermion systems in two different ways: (i) Strong correlations can lead to new topological many-body states that cannot exist without interactions. Fractional topological insulators are an example of such systems~\cite{levin_stern_fractional_TI}.
(ii) Interactions can reduce the classification of free-fermion SPT phases, i.e, two different phases of the free-fermion classification can be continuously connected in the presence of interactions. In that case, we say that the noninteracting classification ``collapses". This possibility was first considered
by Fidkowski and Kitaev\cite{fidkowski2010,fidkowski_kitaev_PRB_11}, who showed that eight Majorana modes localized at the end 
of a one-dimensional topological superconductor with time-reversal symmetry (class BDI) can be gapped out by many-body interactions that are weak relative to the bulk gap. In other words, they found that the $\mathbb{Z}$ classification of one-dimensional superconductors in class BDI reduces to $\mathbb{Z}_8$ when many-body interactions are included~\cite{turner_reduction_entaglement}.

Later, these considerations were generalized to all free-fermion SPT states of the ten-fold way.
In particular, it was shown that the $\mathbb{Z}$ classification of free-fermion systems with chiral symmetry in odd dimensions reduces to $\mathbb{Z}_N$~\cite{morimoto2015,fidkowski2013,metlitski2014,kitaev_kitp,kitaev,you2014,kapustin14_1,kapustin14_2,Kapustin2015,wangscience14,wang2015,tang2012,queiroz_stern_arXiv_16}.
This result was obtained by various different methods, using quantum nonlinear $\sigma$ models (QNLSMs)~\cite{morimoto2015,fidkowski2013,metlitski2014,kitaev_kitp,kitaev,you2014}, cobordism~\cite{kapustin14_1,kapustin14_2,Kapustin2015}, 
vortex condensation~\cite{wangscience14,wang2015}, and group cohomology~\cite{tang2012}. 
These works have lead to a thorough understanding of the classification of interacting SPT states of the ten-fold way. Less is known, however, about the collapse of the classification of free-fermion SPT phases that are protected by crystalline space group symmetries. These space group symmetries are present in any condensed matter system and are, in general, also respected by the interactions.
While this question has been studied for some cases~\cite{ryu2012,yao2013,qi2013,gu2014a,song2016}, no systematic classification of strongly 
correlated SPT states with crystalline symmetries has been obtained so far. 

In this paper, we present a systematic classification of strongly correlated SPT states with order two symmetries that leave the surface invariant, i.e., 
reflection and two-fold rotations.
In particular, we investigate the case where many body interactions lead to a collapse of the classification of free-fermion SPT phases. (The more exotic phases that cannot be adiabatically connected to a free-fermion SPT state are beyond the scope of this paper\cite{wang2014,gu2014b,xu2015,song2016}.)
To derive the reduction pattern  we employ the QNSLM method, in which one considers quartic contact interactions which 
do not break the defining symmetries, neither explicitly nor spontaneously.
The effect of these quartic interactions on the $\nu$ boundary modes is then described by a QNLSM with a target space that depends
on  $\nu$. With this, the collapse of the classification follows from the smallest value of $\nu$ for which the target space has trivial topology.
This approach was first introduced by Kitaev~\cite{kitaev_kitp,kitaev} and later on used
by Morimoto \textit{et al.}~\cite{morimoto2015} to derive the collapse of the ten-fold classification. 
For the case of SPT states with reflection or two-fold rotation symmetry, we find that the noninteracting $\mathbb{Z}_2$ classifications are
stable in the presence of quartic interactions, whereas the $\mathbb{Z}$ classifications are all unstable and
reduce from $\mathbb{Z}$  to $\mathbb{Z}_N$,
where $N$ depends on the spatial dimension $d$ [see Eq.~\eqref{main_reduction_formula}].
These results are summarized in Table~\ref{tab:results} and Table~\ref{tab:results2}.
We illustrate this reduction pattern by a number of physically interesting examples, namely,
a Majorana chain with two-fold rotation symmetry,
a two-dimensional spin-singlet superconductor with time-reversal and reflection symmetry, and 
a class BDI reflection-symmetric topological state (see Sec.~\ref{section_examples}).
For the latter example we show that the classification derived using the QNLSM approach agrees with 
the stability analysis of the surface states using bosonization (Sec.~\ref{examp_BDI_bosnization_analysis}).



The remainder of this paper is organized as follows. 
In Sec.~\ref{sec_strategy}, we briefly review the QNSLM method that we use to study the collapse of the free-fermion classification of SPT states
with reflection and two-fold rotation. We also discuss in this section how the  Hamiltonians of the tenfold way can be represented using either complex fermion or real Majorana operators. It is checked that these two representations give a consistent reduction pattern.
The collapse of the free-fermion classification of SPT states with reflection and rotation symmetry is presented in Sec.~\ref{sec_collapse_reflection}. 
Sec.~\ref{procedure} gives a brief summary of the procedure used to to obtain this result.
In Sec.~\ref{section_examples}, we illustrate the reduction pattern of the classification by considering three examples.
For the case of a three-dimensional topological insulator we show that the reduction pattern obtained by the QNSLM approach is consistent
with a stability analysis of the boundary modes that relies on bosonization techniques.
Our conclusions and outlook are given in Sec.~\ref{sec_conclusion}. Some technical details are relegated to three Appendices.

\section {Symmetries and review of QNLSM approach}
\label{sec_strategy}

In this section we first discuss the symmetry classes in the presence of reflection or two-fold rotation symmetry.
We then give a brief review of  the QNLSM method and explain how the Hamiltonians
can be expressed either with interacting complex fermion or real Majorana operators and discuss some important differences and connections between these two representations.


\subsection{Symmetry classes of crystalline TIs and TSCs}
\label{symmetry_classes_of_TIs_TSCs}

If one disregards crystalline symmetries, all free-fermion systems can be categorized by the ten  Altland-Zirnbauer (AZ) symmetry classes~\cite{altlandZirnbauerPRB97,Zirnbauer:1996fk},
which are distinguished by the presence or absence of 
time-reversal symmetry (TRS), with operator~$\mathcal{T}$, particle-hole symmetry (PHS), with operator $\mathcal{C}$, and chiral symmetry (CHS), with operator $\Gamma$. For a brief review on how these symmetries act on the Hamiltonians, either written in terms of complex fermion operators or real Majorana operators, see Appendix~\ref{appendix_A}.

 An important point to note is that SPT states
of a given AZ symmetry class can be interpreted in different ways.
That is, for a given AZ symmetry class there are different symmetry embedding schemes.
To explain this, let us consider as an example symmetry class BDI.
One-dimensional systems that belong to this symmetry class can be viewed either as Majorana chains with only time-reversal symmetry, or alternatively, as polyactylene chains of complex fermions with time-reversal ($\mathcal{T}^2 = +1$)
and sublattice symmetry. In the latter case one has an additional U(1) symmetry due to charge conservation.
The reduction pattern of the free-fermion classification henceforth might, in principle, depend on 
which interpretation of the AZ symmetry class is used, i.e.,
 which symmetry embedding scheme is used. 
 is also developed using Majorana representation[see Appendix \ref{symmetry} for symmetry operations and Table\ref{tab:symmetries} for symmetry classes], which also yields the same ten symmetry classes. 
%
We find that this is indeed the case for symmetry classes  BDI, DIII and D with reflection/rotation, while different symmetry embeding schemes give the same reduction pattern for classes  CI, CII, and~C with reflection/rotation
(see also the example in Sec.~\ref{3D_example}).


\subsubsection{Reflection symmetry}
\label{sec_reflect_sym}

Let us now discuss how the presence of reflection symmetry leads to a refinement of the ten AZ classes.
Reflection symmetry, with reflection operator $R$, is the invariance of the Hamiltonian under a spatial reflection about a certain reflection plane.
Without loss of generality, we assume that the reflection plane is perpendicular to the $x_1$ axis. Hence, 
reflection symmetry maps 
\begin{equation}
{\boldsymbol x}=(x_1,x_2\cdots x_d)\rightarrow \bar{{\boldsymbol x}}= (-x_1,x_2\cdots x_d)\nonumber
\end{equation}
in $d$ dimensions
 \footnote{Note that in the following we will assume that the surface is invariant under reflection. However, for one-dimensional 
 systems this condition cannot be satisfied and, hence, our approach cannot be applied to one-dimensional reflection SPT states.}.
 Reflection $R$ acts on the second-quantized operators as~\footnote{Summation over repeated indices is assumed throughout the paper.} 
\begin{equation}
\label{eq:reflection}
\hat{R} \, \Psi_i(\boldsymbol x) \, \hat{R}^{-1}= R_{ij}\Psi_j(\bar{\boldsymbol x}) ,
\end{equation}
where $\Psi_i$'s are complex fermion (real Majorana) operators.
The matrix $R$ is unitary (real and symmetric in the Majorana representation).
Due to a phase ambiguity in the definition of the unitary operator $R$, we can assume that $R$ is Hermitian (i.e., $R^2 =1$)~\footnote{This can always be achieved by absorbing a phase factor in the complex fermion operators, provided the symmetry class has a U(1) symmetry corresponding to fermion number conservation. If  $\tilde R^2=-1$ in a symmetry class without U(1) symmetry, we formally take $R=i\tilde R$ (in the complex basis), where $\tilde R$ represents the actual physical symmetry. We should keep in mind, however, that this substitution \emph{alters} the anti/commutation relations of $R$ with the anti-unitary symmetries TRS and PHS. A similar convention is also
used for two-fold rotation symmetries with operator $U$\label{ft1}},
which is in accordance with the conventions used in Refs.~\onlinecite{chiu2013,chiu2015,morimoto2013}.
With this convention the 
algebraic relations between $R$ and the symmetry operators of TRS and PHS (in complex basis) are uniquely defined
and we can organize the symmetry classes of reflection-symmetric TIs (TSCs)  in terms of these relations. We have
\begin{align} 
\label{indicesComAnti}
\Gamma R = \eta_\Gamma R \Gamma,
\quad
\mathcal{T} R = \eta_{\mathcal{T}} R \mathcal{T},
\quad
\mathcal{C} R = \eta_{\mathcal{C}} R \mathcal{C} ,
\end{align}
where the indices $\eta_{\Gamma}$, $\eta_{\mathcal{T}}$, and $\eta_{\mathcal{C}}$ take values $ \pm 1$
 specifying whether $R$ commutes ($+1$) or anticommutes ($-1$) with the corresponding
symmetry operator $\Gamma$, $\mathcal{T}$, or $\mathcal{C}$~\footnote{
Note that, as stated above, the defining symmetries for a given AZ class depend on whether one uses the complex fermion
or the real Majorana representation. Hence, also the algebraic relations between $R$ and the other (anti-unitary) symmetries
depends on which representation one uses.  We note that the 27 cases are defined for symmetries in the complex basis.}.
Hence, in the presence of reflection symmetry $R$ the ten symmetry classes of the tenfold way are enlarged
to 27 symmetry classes, which are labelled by whether $R$ commutes or  anti-commutes
with  $\Gamma$, $\mathcal{T}$, or $\mathcal{C}$.
These 27 symmetry classes are listed in Table.~\ref{tab:results}, labelled
by $R_{\eta_{\mathcal{T}}}$, $R_{\eta_{\Gamma}}$, and $R_{\eta_{\mathcal{C}}}$ 
for the symmetry classes AI, AII, AIII, C, and D, 
and by $R_{\eta_{\mathcal{T}} \eta_{\mathcal{C}}}$ for the chiral symmetry classes BDI, CI, CII, and DIII.

Before we discuss rotation symmetries, let us remark that in systems with charge conservation  
or with $S^z$ spin conservation there
exists an additional symmetry, namely a continuous U(1) symmetry generated by the charge operator $Q$. (This
becomes apparent when one writes the Hamiltonian using real Majorana operators, see Appendix~\ref{symmetry}.)
Hence, one can also consider the algebraic relations between the reflection operator $R$ and the charge $Q$.
To simplify matters, we assume in the following that $R$ commutes with $Q$, i.e., $[Q,R]=0$. (Note, however,
that when $Q$ corresponds to a conserved  $S^z$ spin quantum number, it is possible that $Q$ \emph{anticommutes}
with reflection. But in that case, one can either map the system onto another symmetry class,
or use $R$ to create a unitary on-site symmetry that can be quotient out, see Appendix \ref{connection}.)

\subsubsection{Two-fold rotation symmetry}

Next, we examine the symmetry classes for systems with a two-fold rotation symmetry.
For simplicity we assume that the rotation axis is along the $x_d$ direction.
Hence the rotation symmetry leaves the $x_d$ coordinate invariant, while it flips the
sign of the other $d-1$ spatial coordinates, i.e.,  
\begin{equation}
\boldsymbol{x}=(x_1,x_2\cdots x_{d-1},x_d)\rightarrow \bar{\boldsymbol{x}} = (-x_1,-x_2\cdots,-x_{d-1},x_d).\nonumber
\end{equation}
Two-fold rotation $U$ acts on the second-quantized operators as
\begin{equation}
\label{eq:reflection}
\hat{U} \Psi_i(\boldsymbol x)\hat{U}^{-1}= U_{ij}\Psi_j(\bar{\boldsymbol x}) .
\end{equation}
Similar to the case of reflection symmetry, we assume that the rotation operator $U$ squares to $+1$, i.e., 
 $U^2=1$. With this convention the commutation relations between $U$ and $\mathcal{T}$, $\mathcal{C}$, and $\Gamma$ are uniquely
 defined, which we denote by
 $U_{\eta_{\mathcal{T}}}$, $U_{\eta_{\Gamma}}$, $U_{\eta_{\mathcal{C}}}$, and
$U_{\eta_{\mathcal{T}} \eta_{\mathcal{C}}}$. Just as in the case of rotation symmetric systems,
there is a total of 27 symmetry classes which are listed in Table~\ref{tab:results2}.
(Note that, as in Sec~\ref{sec_reflect_sym}, we assume that $U$ commutes with the U(1) charge~$Q$.)
 

\begin{table*}
\caption{\label{tab:results}  Collapse of the classification of interacting reflection-symmetric topological crystalline superconductors (TCSCs)/topological crystalline insulators (TCIs).  The first column denotes the algebraic relation of the reflection  symmetry $R$ with the protecting symmetries of the AZ classes as explained in the main text. (Here, we impose $R^2=1$.) By comparing with Table VIII of Ref.~\onlinecite{chiu2015}, one finds that the $\mathbb Z$ classifications collapse, while the $\mathbb Z_2$ classifications remain stable. The columns ``Clifford algebra" lists the relevant Clifford algebra encoding all associated matrices in a certain symmetry class with reflection symmetry, written in complex fermion/real Majorana basis, respectively\cite{morimoto2013}. 
We note that the collapse of the classification is given for any spatial dimension $D$, where the relation between $D$ and $n$ is given by  $D=8n+d$, where $d=1, 2, \cdots 8$ and $n=0,1,2, \cdots$. For symmetry classes BDI, D, and DIII, which exhibit two different symmetry embedding schemes, the reduction pattern from $\mathbb Z$ should be further reduced by two if we embed an additional U(1)$\rtimes Z_2^C$ symmetry to the symmetry classes, since these additional symmetry constraints enlarge the root states.} 
\begin{tabular}{c|c|c|cccccccc}
    \hline 
    \hline
   & & &\multicolumn{8}{c}{$D=8n+d, n=0,1,2\cdots$}\\
    \hline
    Ref. & Class & Clifford Algebra& $d=1$ & $d=2$ & $d=3$ & $d=4$ &$d=5$& $d=6$ & $d=7$ &$d=8$\\
    \hline
    $R $ &A&$Cl_{d+2}/Cl_{d+2}$&$\mathbb Z_{2^{4n+2}}$ &0&$\mathbb Z_{2^{4n+3}}$&0&$\mathbb Z_{2^{4n+4}}$ &0&$\mathbb Z_{2^{4n+5}}$&0\\
    $R_+$&AIII& $Cl_{d+3}/Cl_{d+3}$&0&$\mathbb Z_{2^{4n+2}}$&0&$\mathbb Z_{2^{4n+3}}$&0&$\mathbb Z_{2^{4n+4}}$&0&$\mathbb Z_{2^{4n+5}}$\\
    $R_- $ &AIII&$Cl_{d+2}/Cl_{d+2}$&$\mathbb Z_{2^{4n+2}}$ &0&$\mathbb Z_{2^{4n+3}}$&0&$\mathbb Z_{2^{4n+4}}$ &0&$\mathbb Z_{2^{4n+5}}$&0\\
    \hline
    \multirow{8}{*}{$R_{+(+)}$}&AI&$Cl_{2,d+2}/Cl_{2,d+2}$&$\mathbb Z_{2^{4n+2}}$ &0&0&0&$\mathbb Z_{2^{4n+3}}$&0&$\mathbb Z_2$&$\mathbb Z_2$\\
    &BDI&$Cl_{d+1,4}/Cl_{2,d+1}$&$\mathbb Z_2$&$\mathbb Z_{2^{4n+3}}$&0&0&0&$\mathbb Z_{2^{4n+4}}$&0&$\mathbb Z_2$\\
    &D&$Cl_{d,4}/Cl_{2,d}$&$\mathbb Z_2$&$\mathbb Z_2$&$\mathbb Z_{2^{4n+4}}$&0&0&0&$\mathbb Z_{2^{4n+5}}$&0\\
    &DIII&$Cl_{d,5}/Cl_{3,d}$&0&$\mathbb Z_2$&$\mathbb Z_2$&$\mathbb Z_{2^{4n+4}}$&0&0&0&$\mathbb Z_{2^{4n+5}}$ \\
    &AII&$Cl_{4,d}/Cl_{4,d}$&$\mathbb Z_{2^{4n+1}}$&0&$\mathbb Z_2$&$\mathbb Z_2$&$\mathbb Z_{2^{4n+4}}$&0&0&0\\
    &CII&$Cl_{d+3,2}/Cl_{5,d}$&0&$\mathbb Z_{2^{4n+1}}$&0&$\mathbb Z_2$&$\mathbb Z_2$&$\mathbb Z_{2^{4n+4}}$&0&0\\
    &C&$Cl_{2+d,2}/Cl_{d+3,1}$&0&0&$\mathbb Z_{2^{4n+2}}$&0&$\mathbb Z_2$&$\mathbb Z_2$&$\mathbb Z_{2^{4n+5}}$&0\\
    &CI&$Cl_{2+d,3}/Cl_{2,d+3}$&0&0&0&$\mathbb Z_{2^{4n+2}}$ &0&$\mathbb Z_2$&$\mathbb Z_2$&$\mathbb Z_{2^{4n+5}}$ \\
    \hline
   \multirow{8}{*}{$R_{-(-)}$}&AI&$Cl_{1,d+3}/Cl_{1,d+3}$&0&0&$\mathbb Z_{2^{4n+2}}$&0&$\mathbb Z_2$&$\mathbb Z_2$&$\mathbb Z_{2^{4n+5}}$&0\\
    &BDI&$Cl_{2+d,3}/Cl_{1,d+2}$&0&0&0&$\mathbb Z_{2^{4n+3}}$ &0&$\mathbb Z_2$&$\mathbb Z_2$&$\mathbb Z_{2^{4n+6}}$ \\   
   &D&$Cl_{d+1,3}/Cl_{1,d+1}$&$\mathbb Z_{2^{4n+3}}$ &0&0&0&$\mathbb Z_{2^{4n+4}}$&0&$\mathbb Z_2$&$\mathbb Z_2$\\
    &DIII&$Cl_{d+1,4}/Cl_{2,d+1}$&$\mathbb Z_2$&$\mathbb Z_{2^{4n+3}}$&0&0&0&$\mathbb Z_{2^{4n+4}}$&0&$\mathbb Z_2$\\
    &AII&$Cl_{3,d+1}/Cl_{3,d+1}$&$\mathbb Z_2$&$\mathbb Z_2$&$\mathbb Z_{2^{4n+3}}$&0&0&0&$\mathbb Z_{2^{4n+4}}$&0\\
    &CII&$Cl_{d+4,1}/Cl_{4,d+1}$&0&$\mathbb Z_2$&$\mathbb Z_2$&$\mathbb Z_{2^{4n+3}}$&0&0&0&$\mathbb Z_{2^{4n+4}}$ \\
    &C&$Cl_{3+d,1}/Cl_{d+2,2}$&$\mathbb Z_{2^{4n+1}}$&0&$\mathbb Z_2$&$\mathbb Z_2$&$\mathbb Z_{2^{4n+4}}$&0&0&0\\
    &CI&$Cl_{d+3,2}/Cl_{1,d+4}$&0&$\mathbb Z_{2^{4n+1}}$&0&$\mathbb Z_2$&$\mathbb Z_2$&$\mathbb Z_{2^{4n+4}}$&0&0\\    
    \hline
    $R_{-+}$&BDI& $Cl_{d+4}/Cl_{d+2} $&$\mathbb Z_{2^{4n+2}}$&0&$\mathbb Z_{2^{4n+3}}$&0&$\mathbb Z_{2^{4n+4}}$&0&$\mathbb Z_{2^{4n+5}}$&0\\  
     $R_{-+}$&CII& $Cl_{d+4}/Cl_{d+4}$&$\mathbb Z_{2^{4n+1}}$&0&$\mathbb Z_{2^{4n+2}}$&0&$\mathbb Z_{2^{4n+3}}$&0&$\mathbb Z_{2^{4n+4}}$&0\\  
     $R_{+-}$&DIII& $Cl_{d+4}/Cl_{d+2}$&$\mathbb Z_{2^{4n+2}}$&0&$\mathbb Z_{2^{4n+3}}$&0&$\mathbb Z_{2^{4n+4}}$&0&$\mathbb Z_{2^{4n+5}}$&0\\   
     $R_{+-}$&CI& $Cl_{d+4}/Cl_{d+4}$&$\mathbb Z_{2^{4n+1}}$&0&$\mathbb Z_{2^{4n+2}}$&0&$\mathbb Z_{2^{4n+3}}$&0&$\mathbb Z_{2^{4n+4}}$&0\\  
     \hline
     $R_{+-}$&BDI&$Cl_{d+1,3}/Cl_{1,d+1}$&$\mathbb Z_{2^{4n+3}}$&0&0&0&$\mathbb Z_{2^{4n+4}}$&0&$\mathbb Z_2$&$\mathbb Z_2$\\
     $R_{+-}$&CII&$Cl_{d+3,1}/Cl_{4,d}$&$\mathbb Z_{2^{4n+1}}$&0&$\mathbb Z_2$&$\mathbb Z_2$&$\mathbb Z_{2^{4n+4}}$&0&0&0\\
     $R_{-+}$&DIII&$Cl_{d,4}/Cl_{2,d}$&$\mathbb Z_2$&$\mathbb Z_2$&$\mathbb Z_{2^{4n+4}}$&0&0&0&$\mathbb Z_{2^{4n+5}}$&0\\
     $R_{-+}$&CI&$Cl_{2+d,2}/Cl_{1,d+3}$&0&0&$\mathbb Z_{2^{4n+2}}$&0&$\mathbb Z_2$&$\mathbb Z_2$&$\mathbb Z_{2^{4n+5}}$&0\\
     \hline
     \hline
     \end{tabular}
     \end{table*}

\begin{table}
\caption{\label{tab:results2} Collapse of the classification of interacting two-fold rotation-symmetric TCSCs/TCIs.  The first column denotes the commutation relation of the rotation symmetry $U$ with the protecting symmetries of the AZ classes. (Here, we impose $U^2=1$.). Compared with the noninteracting classification\cite{shiozaki2014}, the $\mathbb Z$ classifications collapse, while the $\mathbb Z_2$ classifications remain stable.  We note that the collapse of the classification
is given for any dimension $D=8n+d$, where $d=1,2, \cdots 8$ and $n=0,1,2, \cdots$. For symmetry classes BDI, D, and DIII, that allow for two different symmetry embedding schemes, the reduction pattern from $\mathbb Z$ should be further reduced by two if we embed an additional U(1)$\rtimes Z_2^C$ symmetry to the symmetry classes, since these additional symmetry constraints enlarge the root states.} 
\begin{tabular}{c|c|cccccccc} 
    \hline 
    \hline
   & & \multicolumn{8}{c}{$D=8n+d, n=0,1,2\cdots$}\\
    \hline
    Rot. & Class &  $d=1$ & $d=2$ & $d=3$ & $d=4$ &$d=5$& $d=6$ & $d=7$ &$d=8$\\
    \hline
    $U $ &A&0&0&0&0&0&0&0&0\\
    $U_+$&AIII& $\mathbb Z_{2^{4n+2}}$ &$\mathbb Z_{2^{4n+2}}$&$\mathbb Z_{2^{4n+3}}$&$\mathbb Z_{2^{4n+3}}$&$\mathbb Z_{2^{4n+4}}$ &$\mathbb Z_{2^{4n+4}}$&$\mathbb Z_{2^{4n+5}}$&$\mathbb Z_{2^{4n+5}}$\\
    $U_- $ &AIII&0 &0&0&0&0 &0&0&0\\
    \hline
    \multirow{8}{*}{$U_{+(+)}$}&AI&$\mathbb Z_{2^{4n+2}}$ &0&0&0&$\mathbb Z_{2^{4n+3}}$&0&$\mathbb Z_2$&$\mathbb Z_2$\\
    &BDI&$\mathbb Z_{2^{4n+3}}$&$\mathbb Z_{2^{4n+3}}$&$\mathbb Z_{2^{4n+3}}$&$\mathbb Z_{2^{4n+3}}$&$\mathbb Z_{2^{4n+4}}$&$\mathbb Z_{2^{4n+4}}$&$\mathbb Z_{2^{4n+5}}$&$\mathbb Z_{2^{4n+6}}$\\
    &D&$\mathbb Z_2$&$\mathbb Z_2$&0&0&0&0&0&$\mathbb Z_2$\\
    &DIII&$\mathbb Z_2$&$\mathbb Z_2$&$\mathbb Z_{2^{4n+3}}$&0&0&0&$\mathbb Z_{2^{4n+5}}$&$\mathbb Z_2$ \\
    &AII&0&0&0&0&0&0&0&0\\
    &CII&$\mathbb Z_{2^{4n+1}}$&$\mathbb Z_{2^{4n+1}}$&$\mathbb Z_{2^{4n+2}}$&$\mathbb Z_{2^{4n+3}}$&$\mathbb Z_{2^{4n+4}}$&$\mathbb Z_{2^{4n+4}}$\!&$\mathbb Z_{2^{4n+4}}$&$\mathbb Z_{2^{4n+4}}$\\
    &C&0&0&0&$\mathbb Z_2$&$\mathbb Z_2$&$\mathbb Z_2$&0&0\\
    &CI&0&0&$\mathbb Z_{2^{4n+2}}$ &$\mathbb Z_2$&$\mathbb Z_2$&$\mathbb Z_2$&$\mathbb Z_{2^{4n+4}}$&0 \\
    \hline
   \multirow{8}{*}{$U_{-(-)}$}&AI&0&0&0&0&0&$\mathbb Z_2$&$\mathbb Z_2$&$\mathbb Z_{2}$\\
    &BDI&$\mathbb Z_{2^{4n+2}}$ &0&0&0&$\mathbb Z_{2^{4n+4}}$ &$\mathbb Z_2$&$\mathbb Z_2$&$\mathbb Z_{2}$ \\   
   &D&0&0&0&0&0&0&0&0\\
    &DIII&$\mathbb Z_{2^{4n+2}}$&$\mathbb Z_{2^{4n+3}}$&$\mathbb Z_{2^{4n+4}}$&$\mathbb Z_{2^{4n+4}}$&$\mathbb Z_{2^{4n+4}}$&$\mathbb Z_{2^{4n+4}}$&$\mathbb Z_{2^{4n+5}}$&$\mathbb Z_{2^{4n+5}}$\\
    &AII&0&$\mathbb Z_2$&$\mathbb Z_{2}$&$\mathbb Z_{2}$&0&0&0&0\\
    &CII&$\mathbb Z_{2^{4n+1}}$&$\mathbb Z_2$&$\mathbb Z_2$&$\mathbb Z_2$&$\mathbb Z_{2^{4n+3}}$&0&0&0 \\
    &C&0&0&0&0&0&0&0&0\\
    &CI&$\mathbb Z_{2^{4n+1}}$&$\mathbb Z_{2^{4n+1}}$&$\mathbb Z_{2^{4n+2}}$&$\mathbb Z_{2^{4n+2}}$&$\mathbb Z_{2^{4n+3}}$&$\mathbb Z_{2^{4n+4}}$&$\mathbb Z_{2^{4n+5}}$&$\mathbb Z_{2^{4n+5}}$\\    
    \hline
    $U_{-+}$\!&BDI&$\mathbb Z_{2}$&0&0&0&0&0&$\mathbb Z_2$&$\mathbb Z_2$\\
     $U_{-+}$\!&CII& 0&0&$\mathbb Z_{2}$&$\mathbb Z_2$&$\mathbb Z_{2}$&0&0&0\\  
     $U_{+-}$\!&DIII& 0&0&0&0&0&0&0&0\\   
     $U_{+-}$\!&CI& 0&0&0&0&0&0&0&0\\  
     \hline
     $U_{+-}$\!&BDI&0&0&0&0&0&0&0&0\\
     $U_{+-}$\!&CII&0&0&0&0&0&0&0&0\\
     $U_{-+}$\!&DIII&$\mathbb Z_2$&$\mathbb Z_2$&$\mathbb Z_{2}$&0&0&0&0&0\\
     $U_{-+}$\!&CI&0&0&0&0&$\mathbb Z_2$&$\mathbb Z_2$&$\mathbb Z_{2}$&0\\
     \hline
     \hline
     \end{tabular}
     \end{table}

\subsection {QNLSM approach}
\label{sec_QNLSM_approach_review}

Let us now describe the details of the QNLSM approach~\cite{kitaev_kitp,kitaev,morimoto2015} that we use to derive the reduction pattern of the free-fermion classification. 
The basic idea behind this approach is to study whether the boundary modes of an SPT state with a given set of symmetries can be
gapped out by symmetry-preserving interactions that are weak relative to the bulk gap. Hence, as a first step, we need
to derive the surface Hamiltonian describing the dynamics of the boundary modes. To that end, we start from a family of Dirac
Hamiltonians  representing crystalline SPT states of fermions in $d$ spatial dimensions 
\begin{equation}
\label{eq:hamiltonian}
\begin{split}
\mathcal{H}^{(0)}=-i\sum_{j=1}^d\frac{\partial}{\partial x^j}\tilde \gamma_j\otimes \mathds{1} +m({\boldsymbol x}) \, \tilde\beta\otimes \mathds{1} .
\end{split}
\end{equation}
Here, $\tilde\gamma_j$ and $\tilde\beta$ are anti-commuting Dirac matrices and $\mathds{1}$ is the unit matrix of rank $\nu \in \mathbb{Z}^+$ (the precise meaning of $\nu$ will be explained below).
We choose the rank $r$ of the matrices $\tilde\gamma_j$ and $\tilde\beta$ to be the minimal dimension $r_{\textrm{min}}$ 
which is needed to implement the defining symmetries of the crystalline SPT state. In the following, we call the Hamiltonian 
$\mathcal{H}^{(0)}$ with $\nu=1$ the ``root state" of the corresponding symmetry class. Mathematically speaking, 
the root state is the generator of the Abelian group $\mathcal{B}$, which indexes the different equivalence classes of SPT states
for a given set of symmetries. 
With this choice of $r$, the dimension $\nu$ of the unity matrix $\mathds{1}$ in Eq.~\eqref{eq:hamiltonian} corresponds to the number of copies of root states that we use to test the stability of the boundary modes against interactions
\footnote{Note that the minimal rank $r_{\textrm{min}}$ can, in principle,
depend on whether one uses complex fermion or real Majorana operators to express the Hamiltonian.
But the number of fermion flavors in the root state (i.e, the dimension of the Fock space) does not depend
on whether complex fermion or real Majorana operators are used.}.




Let us now determine the  surface Hamiltonian of Eq.~\eqref{eq:hamiltonian} for the surface that is perpendicular to the 
$x_d$ direction. This surface is left invariant by the reflection (or rotation) symmetry, and thus exhibits boundary modes
protected by the crystalline (and non-spatial) symmetries. The boundary Hamiltonian can be derived  
by considering a domain wall configuration in the mass term $m({\boldsymbol x})$ along the $x_d$ direction~\cite {jackiw}. 
One finds that the Hamiltonian describing the boundary modes with quartic contact interactions is given by
\footnote{In principle, one could consider more complicated interactions than the ones in Eq.~\eqref{eq:boundary} that might
lead to a further collapse of the free-fermion classification. Hence, strictly speaking, the reduction patterns
of Tables~\ref{tab:results} and~\ref{tab:results2} only give an upper bound for the reduction of the classification.} 
\begin{subequations} \label{eq:boundary}
\begin{eqnarray}
H_{bd}  
&=&
H^{(0)}_{bd}+H^{(int)}_{bd} ,\\
H^{(0)}_{bd} \label{bd_kinetic_term}
&=&
\int d^{d-1}{\boldsymbol x} \, \Psi^\dagger (-i\sum_{j=1}^{d-1}\frac{\partial}{\partial x^j}\gamma_i\otimes \mathds{1})\Psi , \\
H^{(int)}_{bd} \label{eq_interaction_term}
&=& 
\lambda \sum _{\{\beta\}} \int d^{d-1}{\boldsymbol x} \,
[\Psi^\dagger\beta\Psi]^2 ,
\end{eqnarray}
\end{subequations}
where $\Psi$  ($\Psi^{\dag})$ represents either complex fermion or real Majorana annihilation (creation) operators 
(depending on the chosen representation) describing the boundary modes.
The Dirac matrices  $\gamma_i \otimes  \mathds{1}$ have dimension $\nu \, ( r_{\textrm{min}} / 2) $ and are obtained
by projecting the matrices $\tilde{\gamma}_i \otimes  \mathds{1}$ in Eq.~\eqref{eq:hamiltonian} onto the surface.
The interaction strength $\lambda$  
is assumed to be independent of $\beta$ and to be positive corresponding to repulsive interactions. 
In order to gap out the boundary modes within a mean-field approximation, the boundary mass matrices $\beta$ 
in the interaction term~\eqref{eq_interaction_term} must  be chosen to anticommute with the Dirac matrices 
$\gamma_i$. In addition, we assume that $\{ \beta \}$ is a pairwise anticommuting set of  matrices.
We note that, if the SPT state is topologically non-trivial
in the free-fermion limit, then
the fermion (Majorana) bilinear 
$\Psi^\dagger\beta\Psi$ has to break at least one of the defining symmetries. 
  
%

Now we can decompose the quartic interaction~\eqref{eq_interaction_term} 
using Euclidean time path integrals and a Hubbard-Stratonovich transformation
with respect to the bosonic fields $\phi_\beta$ conjugate to the bilinear $\Psi^\dagger\beta\Psi$.
This yields a dynamical boundary Hamiltonian which is quadratic in the fermion (Majorana) operators
\begin{eqnarray} \label{dynamical_boundary_ham}
H^{(dyn)}_{bd}(\tau,{\boldsymbol x} ) &= \widetilde{H}^{(0)}_{bd}({\boldsymbol x})+\sum\limits_{\{\beta \} }2i\, \beta \, \phi_\beta(\tau, {\boldsymbol x}) ,
\end{eqnarray}
with the imaginary time $\tau$ and the Lagrangian
\begin{eqnarray} \label{lagrangian_bd}
\mathcal L_{bd} &= \Psi^\dagger[\partial_{\tau}+H_{bd}^{(dyn)}]\Psi+\frac{1}{\lambda}\sum\limits_{\beta}\phi_\beta^2 ,
\end{eqnarray}
where $\widetilde{H}^{(0)}_{bd} = (-i\sum_{j=1}^{d-1}\frac{\partial}{\partial x^j}\gamma_i\otimes \mathds{1})$ is the 
free part of the Hamiltonian~\eqref{eq:boundary}.
We observe that, within a saddle-point approximation, the amplitude fluctuations of the vector $\boldsymbol{\phi}$  with the components $\phi_\beta$ are suppressed by the second term in Eq.~\eqref{lagrangian_bd}.
Since the dynamical mass matrices $\beta$ [we also call it Dirac mass] are mutually anticommuting,
the direction of $\boldsymbol{\phi}$ within the mean-field approximation is arbitrary. Hence, 
after rescaling the length of the vector $\boldsymbol{\phi}$ to one,
the mean-field configuration  of $\boldsymbol{\phi}$ forms a ($N(\nu)-1$)-dimensional sphere $S^{N(\nu)-1}$, where $N(\nu)$ is the number of anticommuting boundary mass matrices $\beta$, which 
depends on $\nu$, the chosen number of root states.
Therefore the direction of $\boldsymbol{\phi}$ is chosen by spontaneous symmetry breaking 
with  $N(\nu)-1$ associated Goldstone modes.

The low-energy effective theory describing the fluctuations of these Goldstone modes is given in terms of a QNLSM, which is obtained by use of a gradient expansion
and by integrating out the fermionic fields.
The partition function for this QNLSM reads~\cite{morimoto2015}
\begin{subequations} \label{eq:nlsm}
\begin {equation}
Z_{bd}\approx \int \mathcal D[ \boldsymbol{\phi} ]\delta( \boldsymbol{\phi}^2-1)e^{-S_{\textrm{QNLSM}}-S_{\textrm{top}}},
\end{equation}
where $S_{\textrm{top}}$ is a topological term and $S_{\textrm{QNLSM}}$ is the Euclidian action 
 \begin{equation}
 S_{\textrm{QNLSM}}=\frac{1}{2g}\int d\tau\int d^{d-1}\boldsymbol{x} \, (\partial_i \boldsymbol{\phi} )^2 ,
\end{equation} 
\end{subequations}
with base space $\mathbb{R}^{(d-1)+1}$ and target space $S^{N(\nu)-1}$.
The topological term $S_{\textrm{top}}$ can only be present in the QNLSM, if any one of the
homotopy groups $\pi_\iota \left[ S^{N(\nu)-1} \right]$, with $\iota = 0, 1, \ldots, d+1$, is nonvanishing~\cite{Abanov2000685}.
The presence of a topological term in the QNLSM~\eqref{eq:nlsm} signals the existence of zero modes of the Hamiltonain~\eqref{dynamical_boundary_ham}
that are localized at topological defects in the order parameter $\boldsymbol{\phi}$.
These zero-modes, in turn, prevent the interactions from
gapping out the boundary modes of the SPT state.
It follows that $\nu$ copies of the root state of an interacting SPT phase
cannot be connected to a trivial state, whenever $\pi_\iota \left[ S^{N(\nu)-1} \right]$ is non-zero for some $\iota$.

On the other hand, if
\begin{eqnarray}
\pi_\iota \left[ S^{N(\nu)-1} \right]  = 0, \quad \textrm{for all $\iota = 0, 1, \ldots, d+1$},
\end{eqnarray}
there is no topological term in the QNLSM. We denote the smallest value of $\nu$ for which this happens by $\nu_{\textrm{min}}$.
By computing the homotopy groups of the spheres, one finds that $\nu_{\textrm{min}}$ must satisfy
the condition
\begin{equation}
\label{eq:criterion}
d+1<N(\nu_{min})-1 .
\end{equation}
In the absence of a topological obstruction, Eq.~\eqref{eq:nlsm} is simply a QNLSM on the sphere
$S^{N(\nu_{\textrm{min}})-1}$. In that case the strong coupling fixed point $g\rightarrow \infty$
of the QNLSM is stable, which corresponds to a quantum-disordered phase
in which all the  the discrete $Z_2$ symmetries are
dynamically restored by quantum fluctuations. 
In order to check that this strong-coupling phase is also compatible with the
continuous symmetries (e.g., a U(1) symmetry corresponding to fermion number conservation),
one needs to verify that the Hubbard-Stratonovich fields 
$\phi_\beta$ are invariant as a set under conjugation with the generators of the continuous symmetries.
That is, the QNLSM target space $S^{N(\nu_{min})-1}$ must remain invariant under the continuous symmetry operations.
If all of these conditions are satisfied, then there exists a continuous symmetry-preserving deformation
that connects  $\nu$ copies of the root state  to a trivial SPT state. Hence, the free-fermion classification
is reduced from, e.g., $\mathbb{Z}$ to $\mathbb{Z}_{\nu_{\textrm{min}}}$.

In closing this section, we remark that there exists an interesting connection between interacting fermionic SPT states and bosonic SPT states
with the same symmetries. That is, the QNLSM~\eqref{eq:nlsm} in $d-1$ spatial dimensions with $N(\nu)=d+2$ bosonic fields $\boldsymbol{\phi}$
and a WZ topological term can be viewed as an $O(d+2)$ nonlinear $\sigma$ model describing the boundary of a $d$-dimensional 
bosonic SPT phase~\cite{bi2015,you2015,vishwanath2013,you2014}. Using this connection, the classification of bosonic SPT states can be
inferred from their interacting fermionic counterparts.




 
\subsection{Complex fermion vs.\  real Majorana representation}
\label{two basis}

As stated above, the reduction patterns of the free-fermion classifications can be derived by expressing the Hamiltonians of the SPT states using either complex fermion~\cite{morimoto2015} or real Majorana operators~\cite{fidkowski2013}.
Both choices give consistent reduction patterns, which we demonstrate in Appendix~\ref{connection}.
In the main text of this paper, however, we focus on the real Majorana representation,
since in this representation the continuous U(1) symmetries are realized explicitly.


But before proceeding, let us briefly highlight the crucial differences between the two representations.
Using the Majorana representation, the root state for a given symmetry class is written 
as
\begin{equation} \label{ham_majorana_rep}
\mathcal{H}^{(0)}
=
- i\chi_a \left[ \sum_{j=1}^d \left(  \frac{\partial}{\partial x^j}   \tilde\gamma_j \right)_{ab}+m ( {\boldsymbol{x}} ) \, \tilde\beta_{ab} \right] \chi_b ,
\end{equation}
where $\chi_a$ are Majorana fields which are related to the fermion operators $\psi_j$ via
$\chi_{2 j -1} = \frac{1}{2} \left( \psi_j + \psi_j^{\dag} \right)$
and 
$\chi_{2 j} = \frac{1}{2 i} \left( \psi_j - \psi_j^{\dag} \right)$.
The matrices $\tilde\gamma_i$ of the kinetic term in Eq.~\eqref{ham_majorana_rep}
 are real symmetric matrices, which
 satisfy $\{\tilde\gamma_i,\tilde\gamma_j\}=2\delta_{ij}$. They all anti-commute with the real anti-symmetric mass matrix $\tilde\beta$. 
 This is in contrast to the complex fermion representation, in which the matrices
 of both the kinetic and mass terms are Hermitian, but not necessarily real and symmetric.

%

Another difference is that the defining symmetries of a given AZ symmetry class depend on whether one uses the complex fermion or the real Majorana representation, which is summarized in Table~\ref{tab:symmetries}. 
For example, continuous U(1) symmetries (due to charge or $S^z$ spin conservation)
with generator $Q$ are  realized trivially in the complex fermion representation,
namely, as $\psi_j \rightarrow e^{i\theta}\psi_j$.  Hence,  if one uses the complex fermion representation,
U(1) symmetries do not belong to the defining symmetries.
In the Majorana representation, on the other hand, the U(1) symmetry is implemented explicitly, as
$\chi_a \rightarrow e^{Q\theta}\chi_a$, with 
$Q$ a real anti-symmetric matrix satisfying $Q^2=-1$.
This difference between complex and real Majorana representations results in ambiguities for the  interpretation of the symmetry classes, cf.~Table~\ref{tab:symmetries}.

A further point to note is that the rank of the Dirac matrices in the root state can be different in the two
representations. (The dimension of the Fock space, however, is the same, see Appendix. \ref{connection}.)
That is, in the presence of a continuous U(1) symmetry with charge $Q$, the rank of the Dirac matrices
in the complex fermion representation is half as big as in the real Majorana representation, since 
the U(1) symmetry can be realized in a trivial way in the complex fermion basis.
Implementing the U(1) symmetry trivially, however, is problematic if one wants to include ``superconducting fluctuations", i.e., Dirac
masses that break the U(1) symmetry. In that case one needs to re-enlarge the rank of the matrices
by introducing a particle-hole grading~%
\footnote{
To include superconducting fluctuations induced by interactions, it is
necessary to re-enlarge the dimension of the root state by adding a particle-hole grading\cite {chiu2015}
\begin{equation}
\label{eq:bdg}
\begin{split}
& \Psi\rightarrow(\psi_1,\psi_2\cdots \psi_n,\psi_1^\dagger,\psi_2^\dagger\cdots \psi_n^\dagger)^T\\
& \mathcal H^{(0)}\rightarrow \mathcal H_{BdG}^{(0)}=[\mathcal H^{(0)}\otimes \mathds 1]\oplus[-\mathcal H^{(0)*}\otimes \mathds 1] ,
\end{split}
\end{equation}
where the U(1) generator becomes $\mathds{1}\otimes \sigma_3$ with $\sigma_3$ a Pauli matrix acting in the particle-hole space.}. 
Thus, using the complex fermion representation leads
to unnecessary complications, and we will therefore put it aside for now.
\begin{table*}[ht]
\caption {\label{tab:symmetries}
The protecting symmetries of the ten AZ symmetry classes in complex fermion and real Majorana fermion rerpresentation. For symmetries in the Majorana rerpresenation the $\pm 1$ in ``$\mathcal T(\pm1)$"  denotes the square of TRS. $r_{com}$ and $r_{real}$ denote the rank of the root state Hamiltonian written in complex and real Majorana representation, respectively. SU($2$) spin-rotation symmetry can be viewed as the three continuous symmetries $e^{Q\theta}$, $e^{C\theta}$, and $e^{QC\theta}$, with $\{Q,C\}=0$. Hence,
SU($2$)  symmetry corresponds to a U(1) symmetry and PHS $\mathcal C$ in the AZ classes.
The last column lists the relation between the  root state rank in complex fermion   and
real Majorana representation. For the AZ classes BDI, D, DIII , CI, C, and CII the last column
also indicates the differences between different symmetry embedding schemes.  (For the classes BDI, D, and DIII these differences arise depending
on whether or not one implements an additional U(1) symmetry. For the classes C, CI, and CII there are different possibilities regarding the algebraic
relations between $\mathcal{T}$ and the generators of the continuous symmetries, see Appendix~\ref{appendix_A} and~\ref{clifford}.)
 The semiproduct~$\rtimes$ implies that elements of the two symmetry groups do not commute.}
\begin{tabular}{c|ccc|c|c}
  \hline 
  \hline
  class\! &\multicolumn{3}{c|}{AZ classes} &\multicolumn{1}{c|}{Majorana basis} & explanation \\
  \hline
               &$\tilde {\mathcal T}$& $\mathcal C$ &$\Gamma$ &  symmetries    &                        \\
               \hline
          A   &   0          & 0  &  0               & U(1)             &  \multirow {2}{0.55\linewidth}{$r_{com}=r_{real}/2$ by virtue of the U(1) symmetry. For AIII, the chiral symmetry $\Gamma$ is time-reversal in Majorana basis.}\\
          AIII&  0          &  0  &  1              &U(1)$\times \mathcal T(+1)$& \\
       \hline
     AI &  1  &0  & 0   &U(1)$\rtimes \mathcal T(+1) $&  $r_{com}=r_{real}/2$ \\
     \hline
   BDI & 1 & 1 & 1 &   $\mathcal T(+1)$&\multirow{3}{0.55\linewidth}{$r_{com}=r_{real}$ (Nambu spinors in complex basis). Physical $R/U$ \emph{always} commutes with ``built-in" PHS $\mathcal C$. For the symmetry embedding scheme with U(1)$\rtimes [Z_2^C\times \mathcal T]$($C^2=1$, Majorana basis scenario (iv) in App. \ref{clifford}) $r_{real}$ doubles.}\\
   D   & 0 & 1 & 0 &  no sym.  &\\
   DIII &-1&1 &1  & $\mathcal T(-1) $ &\\
   \hline
   AII & -1 &0 &0 & U(1)$ \rtimes \mathcal T(-1)$ &  $r_{com}=r_{real}/2$ \\
   \hline
   CII & -1 & -1 &1 & SU($2$)$\times \mathcal T(+1)$& \multirow {3}{0.55\linewidth}{$\tilde {\mathcal T}=\mathcal TC$ (in Majorana basis). $r_{com}=r_{real}/2$. The symmetry embedding scheme with U(1)$\rtimes [Z_2^C \times \mathcal T](\tilde {\mathcal T}=\mathcal T,C^2=-1)$ (Majorana basis scenario (iv) in App. \ref{clifford}) yields the same Clifford algebra. }\\
   C & 0 &-1 & 0 & SU($2$)  & \\
   CI & 1 &-1 & 1&  SU($2$)$\times \mathcal T(-1)$ &  \\
   \hline
   \end{tabular}
   \end{table*}
\section{Reduction of the free-fermion classification of SPT states with reflection and rotation symmetry}
\label{sec_collapse_reflection}

From the strategy described in Sec.~\ref{sec_QNLSM_approach_review}, 
it becomes apparent that the main task in deriving the reduction patterns is
to determine the largest possible QNLSM target space $S^{N(\nu)-1}$  for each value of $\nu$. (Here, $\nu$ is the chosen 
number of root states.) $N(\nu)$ is determined by the largest number 
of symmetry allowed anti-commuting mass matrices $\beta$.
Therefore, we need to study 
 the space of the normalized dynamical boundary mass matrices (Dirac masses),
which is determined, in parts, by the classifying space of an extension problem of Clifford algebras~\cite{kitaev2009,morimoto2013,Stone:2011qo}.

Before proceeding with deriving the reduction patterns, we first review
some basics facts about Clifford algebras, their extensions, 
and how these are related to 
the classification problem of free-fermion SPT states.

\subsection{Clifford algebras and their extensions}
\label{cliff_algebras}

In the following we consider complex as well as real Clifford algebras, which are associative algebras
with generators that anti-commute with each other.
A complex Clifford algebra $Cl_n$ has $n$ generators $e_i$ (complex Hermitian matrices) satisfying 
\begin{equation} \label{def_complex_Cliff}
\{e_a ,e_b \}=2\delta_{a, b}.   
\end{equation}
The products $e_1^{p_1} e_2^{p_2} \cdots e_n^{p_n}$ ($p_i = 0,1$) with complex coefficients
form a $2^n$-dimensional complex vector space.

A real Clifford algebra $Cl_{p,q}$ has $p+q$ generators $e_i$ ($p$ anti-symmetric real matrices and $q$ symmetric real matrices) satisfying 
\begin{eqnarray}
\{e_i,e_j\} &=& 0 \quad (i\neq j), \nonumber\\
e_i^2 &=& 
\left\{ 
\begin{array}{l l}
-1  \quad 1\leq i\leq p,   \\
+1 \quad p+1\leq i\leq p+q . 
\end{array} 
\right. 
\end{eqnarray}
Linear combinations of 
their products with real coefficients form a $2^{p+q}$-dimensional real vector space. 

The classification of free-fermion SPT states can be inferred from possible extensions of the above Clifford algebras.
(This is possible using either the complex fermion or the real Majorana representation of the SPT state.)
For a given AZ symmetry class let us consider a Dirac-Hamiltonian representative with flattened spectrum. The
 kinetic matrices of this Dirac Hamiltonian together with the symmetry operators generate a complex
 Clifford algebra $Cl_{n}$ (for classes A and AIII) or a real Clifford algebra $Cl_{p,q}$ (for classes AI, BDI, D, DIII, AII, CII, C, and CI)~ \cite{kitaev2009,morimoto2013, lu2014}. 
The mass matrix of the Dirac Hamiltonian can be used as an extra generator, leading to a bigger Clifford algebra
$Cl_{p+1,q}(Cl_{p,q+1})$ or $Cl_{n+1}$. Hence, the space of the symmetry-preserving mass matrices is determined by the classifying space 
of the Clifford algebra extensions $Cl_{p,q} \rightarrow Cl_{p+1,q}(Cl_{p,q+1})$ or $ Cl_{n} \rightarrow Cl_{n+1}$.
The classifying spaces for these Clifford algebra extensions are given by
\begin{eqnarray}
&&Cl_{n}\rightarrow Cl_{n+1} \quad \quad \, \text{classifying space} \quad C_{n} , \nonumber\\
&&Cl_{p,q}\rightarrow Cl_{p+1,q} \quad\text{classifying space}\quad R_{p-q+2} , \\
&&Cl_{p,q}\rightarrow Cl_{p,q+1} \quad\text{classifying space}\quad R_{q-p} , \nonumber
\end{eqnarray}
Note that due to Bott periodicity $R_{a+8} = R_a$ and $C_{n+2} = C_n$.
Now, one finds that distinct free-fermion SPT states correspond to topologically distinct extensions of the algebra.
Hence, the free-fermion classification follows from the number of disconnected parts of the classifying spaces
$R_a$ or $C_n$, which corresponds to the number of disconnected parts of the space of the normalized mass matrices. 
This can be computed from the 
zeroth homotopy groups $\pi_0(R_a)$ or $\pi_0(C_n)$, see bottom row of Table~\ref{tab:minimal copies}.

Let us  consider as an example $d$-dimensional SPT states in symmetry class D, 
which have no symmetries when using the Majorana representation.
The relevant Clifford algebra extension problem is $Cl_{0,d}\rightarrow Cl_{1,d}$, generated by
\begin{eqnarray}
\big\{ \tilde\gamma_i, \cdots , \tilde\gamma_d \big\} \;
\rightarrow \;
\big\{ \tilde\gamma_i, \cdots , \tilde\gamma_d , \tilde\beta \big\} .
\end{eqnarray}
The corresponding classifying space is $R_{2-d}$.
Thus the classification of class D SPT states in $d$ dimensions is given by the zeroth homotopy group $\pi_0(R_{2-d})$ 
. 

\subsection {Strategy to determine dynamical boundary mass matrices}
\label{sec_stragetcy_to_find_bdy_mms}

Following the same logic as in Sec.~\ref{cliff_algebras}, we can use Clifford algebra extensions to infer the space of 
the dynamical boundary mass matrices $\beta$. That is, for a given number $\nu$ of root states
we use the classifying space of a Clifford algebra extension to determine the largest number of
anticommuting mass matrices $\beta$ in Eq.~\eqref{dynamical_boundary_ham}, which in turn
gives $N(\nu)$ and, hence, the target space $S^{N(\nu)-1}$ of the QNLSM.

Before proceeding, let us take a moment to re-examine the properties of the dynamical boundary masses. 
First, we note that they are mutually anti-commuting, and that they anti-commute with the kinetic Dirac matrices of the boundary Hamiltonian~\eqref{bd_kinetic_term}. 
Second, we recall from Sec.~\ref{sec_QNLSM_approach_review} that the dynamical boundary masses couple to the bosonic Hubbard-Stranovich field $\boldsymbol{\phi}$, 
which is conjugate to $\Psi^\dagger\beta\Psi$.
Because the strong-coupling phase of the QNLSM must be compatible with the continuous symmetries (e.g., U(1) symmetry), the bosonic fields $\phi_{\beta}$ must be invariant as a set under these symmetries,
which in turn is controlled by the type of the chosen mass matrices $\beta$. In particular, one can, in principle, have a situation where the matrices $\beta$ break the continuous symmetries, but the QNLSM target space $\boldsymbol{\phi}$ remains invariant under the continuous symmetry.
However, if the boundary masses commute with the generators $Q$ of the continuous symmetries,
the QNLSM target space is, of course, automatically symmetric under the U(1) symmetries.


To simplify matters, we will first determine the dynamical boundary masses that are allowed to break all discrete $Z_2$ symmetries, but preserve the continuous symmetries.
If so, one needs to distinguish three different cases: (i) no continuous symmetries (class D), (ii) a U(1) symmetry due to charge or $S^z$ spin conservation (class A), and (iii) an SU(2) symmetry due to spin conservation (class C)~\footnote{The symmetry PHS in the AZ classes CI,CII,C could actually arise from a continuous symmetry, i.e., spin rotation $\chi_a\rightarrow e^{C\theta}_{ab}\chi_b$. In this case, one should treat PHS as a continuous symmetry when searching for allowed Dirac masses.}. In the following we will call these
three cases the ``parent symmetry classes". 
We observe that the algebraic relations of these continuous-symmetry preserving mass matrices with the
kinetic Dirac matrices of the $(d-1)$-dimensional boundary Hamiltonian~\eqref{bd_kinetic_term} are the same as those of the mass matrices of a ($d-1$)-dimensional bulk Hamiltonian in
class D, A, or~C. In other words, the task of finding dynamical boundary masses preserving the continuous symmetries can 
be reduced to the task of finding  
  (extra) mass terms  $\tilde\beta$ of a ($d-1$)-dimensional bulk Hamiltonian [cf.~\eqref{eq:hamiltonian}]~\cite{chiu2013} in class D, A, or C,
see Sec.~\ref{masses_for_parent_classes}.

%

As a second step, we then need to check whether additional dynamical boundary masses can be found that break the continuous symmetries. As shown by detailed calculation\footnote{This is performed by allowing the Dirac masses to break continuous symmetries, i.e., fall into class D and counting $N(\nu)$.}, these continuous-symmetry breaking masses never lead
to a further reduction of the classification. Hence, one can disregard these continuous-symmetry breaking 
masses, and therefore 
the QNLSM target space is always automatically invariant under the continuous symmetries.


%

\subsubsection{Mass matrices for \mbox{the parent symmetry classes D, A, and C}}
\label{masses_for_parent_classes}

\begin{table*}
\caption {
 Minimal number of root state copies $\nu_{\textrm{m}}=2^p$ (where $p$ is listed in the table) for which one can construct a QNLSM without a topological term
 for the three parent symmetry classes $A$, $D$, and $C$, in  $D-1$ spatial dimensions. ($D$ denotes the spatial dimension of the original bulk TCI/TCSC of interest.)
$V_{D-1}$ represents the classifying space in $D-1$ spatial dimensions. The last two lines list the zeroth homotopy group for classifying spaces, which we utilize to arrive at the minimal copy number $\nu_{\textrm{m}}$ for each parent symmetry class. 
}
\label{tab:minimal copies}
\begin{center}
\begin{tabular}{ c|c|cccccccc}
\hline
&&\multicolumn{8}{c}{$D=8n+d,n=0,1,2,3\cdots$}\\
\hline
class& $V_{D-1}$ &$d=1$ & $d=2$ & $d=3$ & $d=4$ &$d=5$& $d=6$ & $d=7$ &$d=8$\\
    \hline
A& $C_{1-D}$&$2+4n$&$2+4n$&$3+4n$&$3+4n$&$4+4n$&$4+4n$&$5+4n$&$5+4n$\\
D& $R_{3-D}$&$2+4n$&$3+4n$&$4+4n$&$4+4n$&$4+4n$&$4+4n$&$5+4n$&$5+4n$\\
C& $R_{7-D}$&$1+4n$&$1+4n$&$2+4n$&$2+4n$&$3+4n$&$4+4n$&$5+4n$&$5+4n$\\
\hline
\multicolumn{2}{c}{$\pi_0 (C_0)\!$ $\pi_0 (C_1)$}& $\pi_0 (R_0)$ & $\pi_0 (R_1)$& $\pi_0 (R_2)$& $\pi_0 (R_3)$& $\pi_0 (R_4)$& $\pi_0 (R_5)$& $\pi_0 (R_6)$&  $\pi_0 (R_7)$\\
\hline
\multicolumn{2}{c}{\!$\mathbb Z$ \quad\quad$0$} &$\mathbb Z$ &$\mathbb Z_2$&$\mathbb Z_2$ &$0$&$\mathbb Z$   &$0$& $0$& $0$\\
\hline
\end{tabular}
\end{center}
\end{table*}

Let us now determine the largest number of anti-commuting boundary mass matrices for the SPT states of
the three parent symmetry classes. To this end, we consider an SPT state of rank $2r$ in $d$ spatial dimensions
which consists of $\nu$ copies of the root state.
Assuming we have already identified one boundary Dirac mass, say $\beta_1$, 
we can view the boundary Hamiltonian of these $d$-dimensional SPT states as a ($d-1$)-dimensional
bulk Hamiltonian of rank $r$ belonging to one of the three parent symmetry classes.
(The existence of at least one boundary Dirac mass for all 27 symmetry classes of reflection and rotation-symmetric
SPT states is proved later, in Sec.~\ref{dynamical_mass_reflection_rotation_SPT}.)
Hence,  the maximal number of dynamical boundary masses can be inferred from
the presence or absence of additional mass terms of the 
$(d-1)$-dimensional bulk system. 
The existence of these additional mass terms (which respect the symmetries of the parent symmetry classes D, A or C)
is obtained from the  Clifford algebra extension problems 
\begin{eqnarray}
Cl_{0,d-1} &\rightarrow& Cl_{1,d-1}, \qquad \textrm{for class D}, \nonumber\\
Cl_{d-1} &\rightarrow& Cl_d,  \qquad  \; \; \, \; \; \textrm{for class A},\\
Cl_{d+1,0} &\rightarrow& Cl_{d+1,1},\qquad\textrm{for class C},
\nonumber
\end{eqnarray}
with the classifying spaces $R_{3-d}$, $C_{d-1}$, and $R_{7-d}$, respectively.
The zeroth homotopy group $\pi_0$ of these classifying spaces
determines the existence of an additional (normalized) mass term.
Namely, if $\pi_0$ is non trivial, there exists no additional mass term, and hence the maximal number of dynamical boundary masses is just one.  (This means that the space of the normalized mass matrix cannot be parametrized in a continuous fashion.)
On the other hand, if $\pi_0$ is zero, there exists an additional Dirac mass matrix, leading to two anticommuting masses $\beta_1$ and $\beta_2$.
(This means that the choice of  
 the normalized mass for the $(d-1)$-dimensional SPT state is not unique in a continuous fashion, i.e, it   can be written as $\cos(\theta)\beta_1+\sin(\theta)\beta_2$, with $\theta\in [0,2\pi)$.)
One can then continue the search for additional mass matrices (with fixed matrix rank $r$)
by considering the extension problems
 \begin{eqnarray}
&& Cl_{1,d-1}\rightarrow Cl_{2,d-1}, \;  Cl_{2,d-1}\rightarrow Cl_{3,d-1}, \;  \textrm{etc.} , \;\; \textrm{for class D}, \nonumber\\
&& Cl_{d}\rightarrow Cl_{d+1}, \;  Cl_{d+1}\rightarrow Cl_{d+2}\;  \textrm{etc.} , \;\;   \textrm{for class A},  \nonumber\\
&&Cl_{d+1,1} \rightarrow Cl_{d+1,2},\; Cl_{d+1,2}\rightarrow Cl_{d+1,3},\; \textrm{etc.},\;\textrm{for class C}, \nonumber
 \end{eqnarray}
until a nontrivial zeroth homotopy group of the corresponding classifying spaces is encountered. This determines the maximal number of Dirac mass matrices $N(\nu)$
that preserve the continuous symmetries of the given parent symmetry class.

From Sec.~\ref{sec_QNLSM_approach_review} it follows that the QNLSM target space for the determined set of Dirac mass matrices
is $S^{N(\nu)-1}$ and, hence, the homotopy groups $\pi_\iota \left[ S^{N(\nu)-1} \right]$ determine whether a topological term is allowed
in the QNLSM. If the topological term is absent, the boundary modes for the $\nu$ copies of the root state are unstable, and thus the 
classification reduces to $\mathbbm{Z}_{N(\nu)}$.
If a topological term in the QNLSM is still allowed for the determined set of Dirac masses, we
 need to multiply the number of root states by two (i.e., $\nu \to 2 \nu$, and thus
the rank of the boundary Hamiltonian increases from $r$ to $2r$)
 \footnote{Actually the allowed number of Dirac masses could only change upon \emph{doubling} the copy number. The number of allowed mass terms is guaranteed to increase upon doubling the copy number since we could always tensor product one of the Dirac masses with $\sigma_1,\sigma_3$ ($\sigma_i$'s denote the corresponding Pauli matrices), respectively, to generate two new mass matrices and enlarge the set of Dirac masses.} 
and check whether this enlarged Hamiltonian can have more Dirac masses. 
The maximal number of Dirac masses for this enlarged Hamiltonian are obtained, as before, 
from the zeroth homotopy groups of the corresponding classifying spaces.
If the QNLSM for this enlarged Hamiltonian with $2 \nu$ root states still has a topological term (topological obstruction),
one needs to double the number of root states once more, i.e., $2 \nu \to 4 \nu$ , and continue this process until the number of Dirac masses $N(\nu)$ is equal (or larger)
than $d+3$, see Eq.~\eqref{eq:criterion}. 
%
 %

In summary, to determine the largest target space for a given $\nu$, we need to 
 count the number of nontrivial homotopy groups in the sequence
\begin{eqnarray}
&&\pi_0(R_{3-d}/C_{1-d}/R_{7-d}), \nonumber\\
&&\quad\pi_0(R_{4-d}/C_{2-d}/R_{8-d}), \nonumber\\
&&\quad\quad\cdots\\
&&\quad\quad\quad\pi_0(R_{4}/C_{2}/R_{8}), \nonumber
\end{eqnarray}
for the parent symmetry classes D, A, and C, respectively. 
From this follows the minimal number of root state copies $\nu_{\textrm{min}}$ for which one
can construct
a QNLSM without a topological term, see Table \ref{tab:minimal copies}.
This in turn determines the reduction of the classification, i.e., $\mathbbm{Z}_N \to  \mathbbm{Z}_{N(\nu_{\textrm{min}})}$. 


\subsubsection {Dynamical boundary masses for reflection and rotation-symmetric SPT states}
\label{dynamical_mass_reflection_rotation_SPT}

As stated above, there exists at least
one boundary Dirac mass for all 27 symmetry classes of reflection and rotation-symmetric
SPT states. This is the key assumption that we used in the previous section to determine the maximal number
of  dynamical mass matrices for the three parent symmetry classes.
In this subsection we prove that this assumption is indeed correct.
We perform the proof using 
the real Majorana representation of the SPT states. 
Before proceeding with the proof, it is important to recall that the
 dynamical Dirac masses must anti-commute with all the kinetic matrices  of the boundary Hamiltonian and commute with the generators of the continuous symmetries.
For example, for an SU($2$) spin-rotation symmetric system, the masses must commute with the generators $Q$ and $C$, where
 $e^{Q\theta}$, $e^{C\theta}$, and $e^{QC\theta}$ (with  $\{Q,C\}=0$) form the three continuous symmetries of SU($2$).

\paragraph{Reflection-symmetric SPT states}
Reflection symmetry $R$ ($R_x$ for example) acts on Hamiltonian~\eqref{eq:hamiltonian}, written in reciprocal space, as
\begin{equation}
R^{-1}\mathcal H^{(0)}(k_x,k_y,\cdots)R = \mathcal H^{(0)}(-k_x,k_y,\cdots),
\end{equation}
which implies that $\{R , \tilde\gamma_x\}=0$,
$ [R ,\tilde\gamma_j ]=0$
for $j \neq x$, and
$[R ,\tilde\beta]=0$.
For reflection-symmetric SPT states with spatial dimension $d>1$, we 
derive the boundary Hamiltonian by considering a domain wall configuration in the mass term along the direction that is perpendicular to the reflection symmetry direction (i.e., the $x$ direction),  i.e. $m(\mathbf x)=m_0sgn(x_d)$.
The  boundary Hamiltonian describing the edge modes possesses all protecting AZ symmetries together with  rotation symmetry $R_{bd}$,
 the projection of the reflection operator $R$ onto the boundary space.

We now construct the boundary Dirac masses for this boundary Hamiltonian. 
In the following,  $C$ denotes one of the generators of the SU($2$) spin-rotation symmetry, cf.~caption of Table~\ref{tab:symmetries}.
We distinguish between four different cases:

\vspace{-0.45cm}

\begin{subequations} \label{bd_masses_reflection}
\subparagraph{(i)}
 $R$ commutes with $C$ (if it exists) and $R_{bd}^2=+1$.  ---
 In this case, one verifies that
\begin{equation}
\beta=\gamma_xR_{bd}   
\end{equation}
satisfies all   algebraic relations that the Dirac mass term must obey.
Here, $\gamma_x$ denotes the kinetic Dirac matrix of the $x$-direction, projected onto the boundary space.

\vspace{-0.45cm}

\subparagraph{(ii)}
 $R$ commutes with $C$ and $R_{bd}^2=-1$. ---
In this case, we find that the mass term is given by
 \begin{equation}
\beta = \gamma_xR_{bd}\otimes i\sigma_2 ,
\end{equation}
which is an anti-symmetric  mass term  in the Majorana representation.

\vspace{-0.45cm}

\subparagraph{(iii)}
$R$ anticommutes with $C$ and $R_{bd}^2=+1$. --- In this case, the mass matrix is
\begin{equation}
\beta=\gamma_x R_{bd} Q\otimes i\sigma_2
\end{equation} ($Q^2=-1$. So in order for $\beta^2=-1$ we have to tensor product with $i\sigma_2$.)

\vspace{-0.45cm}

\subparagraph{(iv)}
$R$ anticommutes with $C$ and $R_{bd}^2= -1$. --- In this case, the mass matrix is
\begin{equation}
\beta=\gamma_x R_{bd} Q .
\end{equation}
\end{subequations}
One verifies that with the above choices the mass terms satisfy all necessary conditions.


\paragraph{ Rotation-symmetric SPT states}

 Two-fold rotation symmetry $U$ acts on Hamiltonian~\eqref{eq:hamiltonian} as
\begin{equation}
U^{-1}\mathcal H^{(0)}(k_1,k_2,\cdots, k_d)U= \mathcal H^{(0)}(-k_1,-k_2,\cdots, k_d) , \nonumber
\end{equation}
from which it follows that
$\{U,\tilde\gamma_i \}=0$ for $ i\neq d$, $[U,\tilde\gamma_d]=0$, and $[U,\tilde\beta]=0$. 
The boundary Hamiltonian is derived by considering
a domain wall along the $x_d$ direction, such that the
boundary Hamiltonian inherits all symmetries of the bulk Hamiltonian, including
the rotation symmetry $U_{bd}$, i.e., the projection of the rotation operator $U$ onto the boundary space.

To construct the boundary mass terms we consider, as before, four different cases:
\vspace{-0.45cm}

\begin{subequations} \label{bd_masses_rotation}
\subparagraph{(i)}
 $U$ commutes with $C$ (if it exists) and $U_{bd}^2=+1$. ---
 In this case the mass term is
 \begin{eqnarray}
 \beta = U_{bd} \otimes i\sigma_2 .
 \end{eqnarray}
\vspace{-0.45cm}

\subparagraph{(ii)}
 $U$ commutes with $C$ and $U_{bd}^2=-1$. ---
 In this situation the mass term is
  \begin{eqnarray}
 \beta =  U_{bd} . 
 \end{eqnarray}
 Here, $U$ alone is enough as a mass term.
 \vspace{-0.45cm}
 
\subparagraph{(iii)}
 $U$ anticommutes with $C$ and $U_{bd}^2=+1$. ---
In this case the mass term is
\begin{eqnarray}  
U_{bd}Q  .
\end{eqnarray}
 
\vspace{-0.45cm}

\subparagraph{(iv)}
 $U$ anticommutes with $C$ and $U_{bd}^2=-1$. ---
 The mass term is
 \begin{eqnarray}
 U_{bd}Q \otimes i\sigma_2  .
\end{eqnarray}
\end{subequations}
%
 With these choices, the mass terms satisfy all necessary symmetry conditions, in particular, they
 anticommute with all $\gamma_i$'s on the boundary. 
 
 
\vspace{0.4cm}

Eqs.~\eqref{bd_masses_reflection} and~\eqref{bd_masses_rotation} prove the existence of boundary Dirac masses for all 27 symmetry classes of 
reflection-symmetric and rotation-symmetric TIs and TSCs.
This means that for any $(d-1)$-dimensional boundary Hamiltonian with reflection (rotation) symmetry, we can always construct a bulk Hamiltonian in the corresponding parent symmetry class in $d-1$ dimensions. This implies that all $\mathbb Z$ classifications of reflection-symmetric and rotation-symmetric TIs and TSCs are unstable
to quartic interactions,
since it is always possible to find enough number of allowed Dirac mass matrices that yield a QNLSM low-energy theory without topological obstructions
(see also discussion in Sec.~\ref{procedure}).
This is an important difference from that of the case without reflection symmetry, where $\mathbb Z $ classifications in even dimensions are stable\cite{morimoto2015}. \footnote{Please note that the Dirac mass matrix we constructed above may not be the allowed matrix with minimal dimension. They just ensure the existence of such terms.}


\subsection {Determining the rank of the root state} 

\begin{table}
\begin{center}
\caption{Periodic table of isomorphisms between irreducible representations of real Clifford algebras $Cl_{p,q}$ and matrix algebras. The symbols $\mathbb R(N)$,
$\mathbb C(N)$, and $\mathbb H(N)$ denote $N\times N$ matrices over $\mathbb R$, $\mathbb C$, and $\mathbb H$, respectively. With this,
the rank of the root state [realized in the Majorana basis, i.e.,  $GL(R)$] follows from the dimension of the matrix
algebras: $\dim \mathbb R(N)=N$, $\dim \mathbb C(N)=2N$, $\dim \mathbb H(N)=4N$. For the case 
where the matrix algebra is a direct sum of two algebras
[denoted as $2\mathbb {R}(N)$,  $2\mathbb {H}(N)$, and  $2\mathbb {C}(N)$],  the ranks of the root state is determined by the dimension of  the subalgebras of these direct sums, since the subalgebras faithfully capture the algebraic relations.
By virtue of the isomorphism $Cl_{p,q+8}\simeq Cl_{p+8,q} \simeq Cl_{p,q}\otimes \mathbb R(16)$, we get the rank of the root state for all real symmetry classes. The rank of the root state of the complex symmetry classes, realized in the Majorana basis, follows from $\dim (Cl_{2m})=\dim (Cl_{2m+1})=2^{m+1}$.}
\label {tab:isomorphism}
\begin{tabular}{c|cccccccc}
\hline
\hline
q $\backslash$ p & $0$& $1$& $2$ &$3$ & $4$ & $5$ & $6$ & $7$\\
\hline
$0$ & $\mathbb R $ & $\mathbb C$ &$\mathbb H$ &$2\mathbb H$ &$\mathbb H(2)$ &$\mathbb C(4)$ &$\mathbb R(8)$ &$2\mathbb R(8)$ \\
$1$ & $2\mathbb R$ &$\mathbb R(2)$ &$\mathbb C(2)$ &$\mathbb H(2)$ &$2\mathbb H(2)$ &$\mathbb H(4)$ &$\mathbb C(8)$ &$\mathbb R(16)$\\
$2$ & $\mathbb R(2)$ &$2\mathbb R(2)$ &$\mathbb R(4)$ &$\mathbb C(4)$ &$\mathbb H(4)$ &$2\mathbb H(4)$ &$\mathbb H(8)$ &$\mathbb H(16)$\\
$3$ & $\mathbb C(2)$ &$\mathbb R(4)$ &$2\mathbb R(4)$ &$\mathbb R(8)$ &$\mathbb C(8)$ &$\mathbb H(8)$ &$2\mathbb H(8)$ &$\mathbb H(16)$\\
$4$ & $\mathbb H(2)$ &$\mathbb C(4)$ &$\mathbb R(8)$ &$2\mathbb R(8)$ &$\mathbb R(16)$ &$\mathbb C(16)$ &$\mathbb H(16)$ &$2\mathbb H(16)$\\
$5$ & $2\mathbb H(2)$ &$\mathbb H(4)$ &$\mathbb C(8)$ &$\mathbb R(16)$ &$2\mathbb R(16)$ &$\mathbb R(32)$ &$\mathbb C(32)$ &$\mathbb H(32)$\\
$6$ & $\mathbb H(4)$ &$2\mathbb H(4)$ &$\mathbb H(8)$ &$\mathbb C(16)$ &$\mathbb R(32)$ &$2\mathbb R(32)$ &$\mathbb R(64)$ &$\mathbb C(64)$\\
$7$ & $\mathbb C(8)$ &$\mathbb H(8)$ &$2\mathbb H(8)$ &$\mathbb H(16)$ &$\mathbb C(32)$ &$\mathbb R(64)$ &$2\mathbb R(64)$ &$\mathbb R(128)$\\
\hline
\hline
\end{tabular}
\end{center}
\end{table}

 Having obtained the dynamical boundary masses, we can add the pieces of the derivation together, to obtain 
the minimal copies of root states needed for each scenario to arrive at a QNLSM without topological obstructions. Since the number of allowed Dirac masses obeying certain symmetries only depend on the \emph{matrix rank} of the boundary Hamiltonian, which is the product of the copy number and the rank of the root state, we only need to determine the size of the root state on the boundary for each case and compare it with that of the corresponding parent classes (D, A or C) in $d-1$ space dimensions. Then one can determine the space of normalized dynamical Dirac mass terms for each copy number of the boundary root state from that of the corresponding parent symmetry class we derived in Sec. \ref{masses_for_parent_classes}.

To determine the rank of the root state we use the isomorphism between irreducible representations of Clifford algebras and matrix algebras, see Table~\ref {tab:isomorphism}. As before we use the real Majorana representation to do this\footnote{The rank for root states in complex basis could be deduced from its relation to Majorana basis listed in Table \ref{tab:symmetries} if needed}.
As discussed in Sec.~\ref{masses_for_parent_classes}, 
for each AZ symmetry class in a given spatial dimension there exists an associated Clifford algebra, which 
is composed of the kinetic and mass Dirac matrices and the symmetry operators of the AZ symmetries. 
For the three parent classes A, D and C in $d-1$ spatial dimensions we found in Sec.~\ref{masses_for_parent_classes}
that the associated Clifford algebras are $Cl_{d-1}$, $Cl_{1,d-1}$, and $Cl_{d+1,1}$, respectively.
Now, we need to incorporate the reflection (rotation) symmetry in the Clifford algebra. 
This is done in Appendix \ref{clifford}, where we derive the Clifford algebras for all 27 symmetry classes
of reflection- and rotation-symmetric TIs and TSCs. 
For reflection-symmetric TIs and TSCs
the corresponding Clifford algebras are listed in the third column of  Table \ref{tab:results}.
Having identified the associated Clifford algebras, we can then infer the
size of the root state  for each of the 27 reflection (rotation) symmetry classes (as well as for the parent symmetry classes) form the
isomorphisms tabulated in Table \ref{tab:isomorphism}.  
 

\subsection {Summary of procedure to obtain the reduction pattern}  
\label{procedure}

To sum up, the derivation of the reduction pattern of the free-fermion classification of crystalline SPT states consists of the following steps:


(1) The first step is to determine  the root state and its rank  $r_{min}$  for a given symmetry class in $d$ spatial dimensions. 
As discussed in Sec.~\ref{sec_QNLSM_approach_review}, the root state is given by the Hamiltonian  $\mathcal{H}^{(0)}$, Eq.~\eqref{eq:hamiltonian}, with $\nu=1$,
i.e., the Hamiltonian with smallest rank that accomodates all the defining symmetries of the crystalline SPT state.
The rank of the boundary Hamiltonian describing the gapless surface modes is then given by $r_{min}/2$.
For each root state there exists an associated Clifford algebra, see Table \ref{tab:results} and  Appendix~\ref{clifford}.
The rank of the root state is obtained by using
 the isomorphism between irreducible representations of Clifford algebras and matrix algebras, see Table~\ref{tab:isomorphism}.

(2) The second step is to 
determine the dynamical boundary masses for this root state that are allowed to break all discrete $Z_2$ symmetries, but should preserve the continuous symmetries.
This task can be reduced to the task of finding (extra) mass terms of a ($d -1$)-dimensional bulk Hamiltonian in the corresponding
parent symmetry class D, A, or C, whose rank we denote by $r_m$.
[For cases with only a U(1) continuous symmetry, the parent symmetry class is A; for cases with SU(2) rotation symmetry, the parent symmetry class is  C; without continuous symmetries, the parent symmetry class is  D,  see Sec.~\ref{masses_for_parent_classes}.] 
Then, one needs to find the minimal number of copies $\nu_m$ for this $(d-1)$-dimensional bulk Hamiltonian in the parent symmetry class D, A, or C, 
for which on can construct a QNLSM without topological obstructions, 
cf.~Table.~\ref{tab:minimal copies}.
From this it follows, that the boundary modes of $\frac{\nu_m r_m}{r_{min}/2}$ copies of the root state of the crystalline SPT state 
can be gapped out by symmetry-preserving interactions.
Hence, we conclude that the free-fermion classification is, at the very least, reduced to 
\begin{eqnarray} \label{main_reduction_formula}
\mathbb Z_{\frac{2\nu_mr_m}{r_{min}}} .
\end{eqnarray}

(3) Finally, we need to check whether additional dynamical boundary masses can be found that break the continuous symmetries (i.e., Dirac masses that belong to class D). This could, in principle, lead
to a further reduction of the classification.
However, as it turns out, these additional continuous-symmetry breaking masses 
do not exist for any of the considered crystalline SPT states.


\vspace{0.2cm}

Following the above three steps, one obtains the interaction-induced collapse of the free-fermion classification of 
reflection-symmetric and rotation-symmetric TIs and TSCs, 
see Tables~\ref{tab:results} and~\ref{tab:results2}.
 Remarkably, we find that all $\mathbb Z_2$ free-fermion classifications  
  are stable against quartic contact interactions, i.e., interactions cannot gap out a single copy of the corresponding root state boundary system.

\section {Examples in 1,2 and 3 space dimensions}
\label{section_examples}

Let us now illustrate the collapse of the classification of free-fermion crystalline SPT states by considering three physical examples.

\subsection {Kitaev Majorana chain with two-fold rotation symmetry}
\label{kitaev_majorana_chain_with_rotation}

The first example is the one-dimensional   Kitaev Majorana chain with a two-fold rotation symmetry.
In the continuum limit and using the Majorana representation~\cite{you2014}, 
 the root state Hamiltonian of this  one-dimensional superconducting wire reads  
\begin{eqnarray} \label{ham_def_kitaev_majorana_chain}
 \mathcal H^{(0)}=i\partial_x X_{10}+mX_{20}, 
\end{eqnarray}
where $X_{ij}=\sigma_i\otimes\sigma_j$ denotes the tensor product of Pauli matrices ($\sigma_1$, $\sigma_2$, $\sigma_3$) and the unit matrix ($\sigma_0$). We will use this notation throughout this entire section. 
Eq.~\eqref{ham_def_kitaev_majorana_chain} satisfies both time-reversal and rotation symmetry with the symmetry operators
\begin{eqnarray}
\mathcal T=\mathcal K X_{30} \quad \textrm{and} \quad \tilde U=iX_{02}, 
 \end{eqnarray}
respectively.
Here, the two-fold rotation  $\tilde U$, which squares to~$-1$, 
is around the axis of the chain.
We note that the dimension of the root state Hamiltonian is enlarged by two compared to the original Kitaev chain model without rotation symmetry.
Hence, Eq.~\eqref{ham_def_kitaev_majorana_chain} can be viewed as two copies of the original Kitaev chain, i.e., a model
with four Majorana flavors in one unit cell that transform as a spin-1/2 object.

To which symmetry class of Table~\ref{tab:results2} does Hamiltonian~\eqref{ham_def_kitaev_majorana_chain} belong to?
The algebraic relations between the symmetry operators are  
$ [\mathcal T, \tilde U]=0$ and 
$[C,\tilde U]=0$, where $C$ denotes the operator of PHS, which is trivial in the real Majorana representation.
(If we use the complex fermion representation of the root state, $C$ becomes a 
nontrivial ``built-in" PHS, once written in Nambu representation, see Eq.~\eqref{sym_ops_examp_I_complex} and Appendix \ref{symmetry}.) 
As discussed in Sec.~\ref{symmetry_classes_of_TIs_TSCs}, the rotation operator needs to
square to $+1$ according to our conventions. Therefore, we need to formally take
 $U=i\tilde U$, which converts the commutation relations into anti-commutation relations.  
 As a consequence, the root state Hamiltonian~\eqref{ham_def_kitaev_majorana_chain} belongs to symmetry class BDI with $U_{--}$ in Table \ref{tab:results2}.
Alternatively, we can write Eq.~\eqref{ham_def_kitaev_majorana_chain} in the complex fermion (Nambu) representation, i.e.,
\begin{eqnarray} \label{ham_def_kitaev_majorana_chainCOMPLEX}
&&\mathcal H^{(0)}=i\partial_x X_{20}+mX_{30}  ,
\end{eqnarray}
in which case the symmetry operators take the form
\begin{eqnarray} \label{sym_ops_examp_I_complex}
&&\mathcal T=\mathcal K , \quad C=\mathcal K X_{10}, \quad \textrm{and} \quad U=X_{02} .
\end{eqnarray}
One verifies that $U$ anticommutes 
with the TRS and PHS operators of 
Eq.~\eqref{sym_ops_examp_I_complex},
thereby confirming that the root state Hamiltonian belongs to class BDI with $U_{--}$.


The Dirac matrices $\tilde{\gamma}_x = X_{10}$ and $\tilde{\beta}=X_{20}$ of the root state Hamiltonian~\eqref{ham_def_kitaev_majorana_chain} together
 with the symmetry operators $\mathcal T$ and $\tilde U$ generate the Clifford algebra $Cl_{3}$, i.e., 
  $\{\tilde\gamma_x, \mathcal T; \tilde\beta\}\otimes \tilde U$  generates $Cl_{3}$. According to the caption of Table~\ref{tab:isomorphism},
this Clifford algebra has dimension four, i.e., $\dim (Cl_{3})=4$, which agrees with the matrix rank of $ \mathcal H^{(0)}$.
That is, the rank of the root state is $r_{\textrm{min}} =4$.
Furthermore, we note that the boundary Hamiltonian of $ \mathcal H^{(0)}$, Eq.~\eqref{ham_def_kitaev_majorana_chain}, falls into class D, since there are no continuous symmetries.
That is, the parent symmetry class is class D. The rank of the root state in zero spatial dimensions  ($d-1=0$) in the parent symmetry class D is $r_m=2$, since the
relevant Clifford algebra is  $Cl_{1,0}$ (cf.~Table~\ref{tab:isomorphism}).
Now, according to Table \ref{tab:minimal copies},  $\nu_m = 2^2=4$ copies of the class D root states in $d-1$ spatial dimensions are needed to gap out the edge modes.  From Eq.~\eqref{main_reduction_formula} it follows that the classification is $ \mathbb Z_{\frac{2\nu_mr_m}{r_{min}}} = \mathbb Z_4$.
 So we need $4$ copies of the Majorana chain~\eqref{ham_def_kitaev_majorana_chain} to gap out all its edge modes and smoothly connect it to the trivial phase,
 cf.~Table~\ref{tab:results2}.

Alternatively, this result can  also be derived by directly analyzing the dynamical boundary Hamiltonian of Eq.~\eqref{ham_def_kitaev_majorana_chain}. 
We will now do this using the complex fermion (Nambu) representation of our example system~\cite{morimoto2015}, i.e., Eq.~\eqref{ham_def_kitaev_majorana_chainCOMPLEX}.
The boundary Hamiltonian of  Eq.~\eqref{ham_def_kitaev_majorana_chainCOMPLEX} is obtained by considering a domain wall configuration 
in the mass term $m X_{30}$.
Adding quartic contact interactions and performing a Hubbard-Stratonovich transformation yields
the dynamical boundary Hamiltonian (cf.~discussion in Sec.~\ref{sec_QNLSM_approach_review})
 \begin{equation}
H^{(dyn)}_{bd} ( \tau)
=M (\tau).
\end{equation}
Since the boundary Hamiltonian has zero spatial dimension, it contains only the dynamical mass term $M (\tau)$, which 
depends on imaginary time $\tau$.
$M (\tau)$ is a $2\nu\times2\nu$ Hermitian matrix, where $\nu$ denotes the number of root state copies.
On the boundary TRS, PHS, and rotation symmetry are represented by
\begin{eqnarray}
 \mathcal T_{ {bd}}= \mathcal{K} X_0   \mathds 1 ,
 \quad    C_{ {bd}} =\mathcal{K} X_0   \mathds 1,
  \quad \textrm{and} \quad U_{ {bd}} =X_2   \mathds 1,
\end{eqnarray} 
respectively, where $\mathds 1$ is the $\nu \times \nu$ unit matrix. Generic quartic contact interactions that respect the BDI symmetries lead to a dynamical mass term $M (\tau)$ in symmetry class D.
Hence, due to PHS the mass term must satisfy $M^\ast (\tau) = - M ( \tau)$. 
(Note that $M ( \tau)$ is allowed to break TRS and rotation symmetry.)
Furthermore, we require that $M(\tau)$ squares to the 
$2\nu\times2\nu$ unit matrix. 
With these conditions, the space of the dynamical mass matrices is topologically equivalent to~\cite{chiu2015}
\begin{equation}
V_\nu=O(2\nu)/U(\nu), 
\end{equation} 
which in the limit $\nu\rightarrow\infty$ corresponds to the classifying space~$R_2$. 

The edge modes of Hamiltonian~\eqref{ham_def_kitaev_majorana_chainCOMPLEX} can be gapped out dynamically, if the
QNLSM for the dynamical masses $M(\tau)$ does not contain a topological term (topological obstruction), cf.~Sec.~\ref{sec_QNLSM_approach_review}.
In order to check whether the QNLSM contains such a topological term, let us now explicitly construct the spaces of the dynamical mass terms $M(\tau)$
for the copy numbers $\nu=1$, $\nu=2$, and $\nu=4$ in the following.
 

\emph{Case $\nu=1$.} ---
For $\nu=1$ the only allowed Dirac mass term is proportional to $X_2$.
(There does not exist any extra mass term since $\pi_0(R_2)=\mathbb Z_2$, cf.~Sec.~\ref{masses_for_parent_classes}).
Hence, the number of anti-commuting mass matrices is $N ( 1) =1$ and therefore the QNLSM target space is $S^{N(1)-1}=S^0$.
Since $\pi_0 ( S^0 ) = \mathbb{Z}_2$, there exists a topological obstruction, which takes the form of a domain wall
in imaginary time, e.g., $\sim \textrm{sgn} ( \tau) X_2$. Due to this domain wall obstruction the edge modes
cannot be gapped out dynamically for $\nu=1$.

\emph{Case $\nu=2$.} --- 
For $\nu=2$, i.e., two copies of the root state~\eqref{ham_def_kitaev_majorana_chainCOMPLEX}, the space of the dynamical Dirac masses is spanned by
\begin{eqnarray}
X_{20}, \quad X_{12}, \quad  \textrm{and} \quad X_{32} .
\end{eqnarray}
That is,  the number of allowed anti-commuting Dirac mass matrices is $N (2) =3$.
(There is no fourth mass term that can be added since  $\pi_0(R_4)=\mathbb Z$.)
Hence, the space of the normalized boundary masses is homeomorphic to $S^2$, i.e., the QNLSM target space is $S^{N(2)-1}=S^2$.
Because $\pi_2 (S^2 ) = \mathbb{Z}$, a Wess-Zumino topological term can be added to the QNLSM.
Due to this WZ topological term, the boundary Hamiltonian for $\nu =2$ remains gapless in the presence of interactions.

\emph{Case $\nu=4$.} --- 
For $\nu=4$ there exist seven anti-commuting Dirac mass matrices, i.e., $N (4) =7$.
There does not exist an  eighth mass matrix since $\pi_0(R_8)=\mathbb Z$.
Hence, the QNLSM target space is $S^{N(4)-1} = S^6$.
Since $\pi_\iota ( S^6 ) = 0$ for $\iota = 0,1,2$, no topological
term can be added to the QNLSM. As a consequence, for $\nu=4$ the edge modes are gapped out dynamically by interactions.
(Note that for the purpose of gapping out the edge modes, one can choose, for example, the four pairwise  anticommuting Dirac masses
$X_{200}$, $X_{320}$, $X_{332}$,  and $X_{102}$.)

Therefore, we conclude that the classification of  Hamiltonian~\eqref{ham_def_kitaev_majorana_chainCOMPLEX} collapses to $\mathbb Z_4$ in the presence of interactions, which agrees with the previous derivation.

\subsection{ Two-dimensional spin-singlet superconductor with time-reversal and reflection symmetry}

As a second example we consider a two-dimensional spin-singlet superconductor with time-reversal and reflection symmetry.
In the Majorana representation the root state Hamiltonian of this superconductor reads
\begin{eqnarray} \label{examp_two_root_state_ham_Majo}
&&\mathcal H^{(0)}=i\partial_x X_{3100}+i\partial_y X_{0202}+mX_{0302}  ,
\end{eqnarray}
where $X_{ijlk}$
denotes the tensor product of four Pauli/identity matrices.
Hamiltonian~\eqref{examp_two_root_state_ham_Majo} is invariant under time-reversal and reflection symmetry $x \to -x$ with
the symmetry operators
\begin{subequations}
\begin{eqnarray}
&&\mathcal T=iX_{2100}\mathcal K  \quad \textrm{and} \quad R_x=X_{2002},
\end{eqnarray}
respectively.
The root state~\eqref{examp_two_root_state_ham_Majo} also satisfies SU($2$) spin-rotation symmetry with the generators
\begin{eqnarray}
  C=iX_{0123}\quad \textrm{and} \quad Q=iX_{0002}  .
\end{eqnarray}
\end{subequations}
Hence, it follows that Hamiltonian~\eqref{examp_two_root_state_ham_Majo} belongs to AZ symmetry class CI, since it is invariant
under SU($2$)$\times \mathcal T$ with $\mathcal T^2=-1$, see Table~\ref{tab:symmetries}.
We infer that the symmetry $\mathcal T$ combined with the symmetry $\mathcal C$ in the Majorana representation
corresponds to the time-reversal symmetry  $\tilde {\mathcal T}$  in the complex fermion representation,
i.e., $\tilde {\mathcal T} = \mathcal TC$, with  $\tilde{\mathcal T}^2 = +1$. 
Since $\{R,\tilde {\mathcal T}\}=\{R,C\}=0$, our example Hamiltonian is in symmetry class CI with $R_{--}$ in Table \ref{tab:results}. 

From Eq.~\eqref{examp_two_root_state_ham_Majo} we find that the rank of the root state is $r_{\textrm{min}}=16$. 
Since the boundary Hamiltonian of Eq.~\eqref{examp_two_root_state_ham_Majo} has a continuous SU($2$) spin-rotation symmetry,
the parent symmetry class that we need to consider is class C. The rank of the ($d-1$)-dimensional (i.e., one-dimensional) root state Hamiltonian
in parent symmetry class C is $r_m = 8$, because the relevant Clifford algebra is $Cl_{3,1}$ and
$\dim Cl_{3,1}=\dim \mathbb H(2)=8$, see Table~\ref{tab:isomorphism}.
We note that for the present example $r_m$ is equal to the rank of the boundary Hamiltonian. 
Using Table~\ref{tab:minimal copies}, we find that for $\nu_m=2^1=2$ copies of the class D root state
in $d-1=1$ spatial dimensions, it is possible to gap out the edge states.
Hence, according to Eq.~\eqref{main_reduction_formula}, the classification is $ \mathbb Z_{\frac{2\nu_mr_m}{r_{min}}} = \mathbb Z_2$.
In other words, the SPT state~\eqref{examp_two_root_state_ham_Majo} forms a $\mathbb Z_2$ group, which is in agreement with Table~\ref{tab:results}.

As in the previous example, we now present an alternative derivation of this result by explicitly constructing the dynamical mass terms
for the boundary Hamiltonian of Eq.~\eqref{examp_two_root_state_ham_Majo}.
The boundary Hamiltonian is derived by considering a domain wall configuration along the $y$ direction
in the mass term $mX_{0302}$ of Eq.~\eqref{examp_two_root_state_ham_Majo}.
After introducing quartic contact interactions and performing a Hubbard-Stratonovich   transformation, we
obtain 
\begin{eqnarray} \label{examp_bdy_ham_CI}
H_{bd}^{(dyn)}=i\partial_x X_{300}\otimes \mathds 1+M(\tau,x) ,
\end{eqnarray}
where $\mathds 1$ is the $\nu \times \nu$ unit matrix and the mass term $M(\tau,x)$
is an anti-symmetric  $8\nu\times 8\nu$ matrix,
with $\nu$ the number of root state copies.
On the boundary, the operations for TRS and reflection symmetry
are represented by
\begin{eqnarray}
\mathcal T_{bd} =iX_{200}\mathcal K \quad \textrm{and} \quad R_{bd,x}=X_{202},
\end{eqnarray}
respectively, and the generators of the continuous SU(2) symmetry read
\begin{eqnarray}
 C_{bd} =iX_{023}\quad  \textrm{and} \quad Q_{bd} =iX_{002} .
\end{eqnarray}
The dynamical mass matrix 
$M(\tau,x)$ anti-commutes with the kinetic term of Eq.~\eqref{examp_bdy_ham_CI},
commutes with the generators of the SU($2$) symmetry (i.e., $[M,Q_{bd}]=[M,C_{bd}]=0$),
and is required to square to unity. (Note that $M ( \tau, x)$ is allowed to break TRS and reflection symmetry.)
Thus, the space of the SU($2$) symmetric boundary matrices $M(\tau,x)$ is
 topologically equivalent to the space 
\begin{equation}
V_\nu= \textrm{Sp}(\nu),
\end{equation}
which in the limit $\nu\rightarrow\infty$ becomes the classifying space $R_5$. 
As in the previous example, we now explicitly construct the dynamical boundary mass terms 
for the copy numbers $\nu=1$ and $\nu=2$.

\emph{Case $\nu=1$.} ---
There are $N (\nu=1)=4$ dynamical mass matrices that are allowed on the boundary, namely, 
\begin{eqnarray} \label{masses_2nd_examp_nu_1}
X_{200}, \quad X_{112}, \quad X_{120}, \quad \textrm{and} \quad X_{132} . 
\end{eqnarray}
(We can add three additional mass matrices since $\pi_0(R_5)=\pi_0(R_6)=\pi_0(R_7)=0$.
There does not exist a fifth mass matrix since $\pi_0(R_8)= \mathbb Z$.)
The space of the dynamical mass matrices is homeomorphic to $S^{N(1)-1} = S^3$. 
Since $\pi_3 ( S^3 ) = \mathbb{Z}$, a WZ topological term can be added to the QNLSM. In the presence of this WZ term, 
the boundary Hamiltonian remains gapless in the presence of interactions. 
In passing we note that the masses $X_{112}$, $X_{120}$, and $X_{132}$ in Eq.~\eqref{masses_2nd_examp_nu_1} satisfy TRS and SU($2$) symmetry, but break reflection symmetry. This means that a two-dimensional class CI superconductor is topologically trivial in the absence of reflection symmetry. 

\emph{Case $\nu=2$.} ---
For  $\nu=2$ we find that there are $N(2)=5$ anti-commuting mass matrices, since $\pi_0(R_9)=\mathbb Z_2$.
Hence the QNLSM target space is  $S^{N(2)-1} = S^4$. Because $\pi_\iota (S^4)=0$, for $\iota =0,1, \ldots, 3$,
no topological term is possible in the QNLSM. As a consequence, for $\nu=2$ the boundary zero modes are gapped out dynamically, 
which confirms that Hamiltonian~\eqref{examp_two_root_state_ham_Majo} is classified as $\mathbb{Z}_2$.

One can check that allowing for SU($2$) symmetry breaking mass terms will not further reduce this classification.

\subsection{Three-dimensional class BDI insulator/superconductor with reflection symmetry}
\label{3D_example}

 The third example is a three-dimensional class BDI topological state with reflection symmetry.
As discussed in  Secs.~\ref{symmetry_classes_of_TIs_TSCs} and~\ref{two basis}, SPT states in AZ class BDI can be
interpreted in two different ways (i.e., there are two different symmetry embedding schemes):
(i) as superconductors with time-reversal symmetry but broken  U(1)  charge symmetry and
(ii) as insulators with U(1) charge symmetry, time-reversal symmetry, and particle-hole symmetry.
In the following we discuss both of these symmetry embedding schemes and show that 
they lead to different reduction patterns of the free-fermion classification.

%

\subsubsection{BDI superconductor with reflection symmetry}
\label{BDI_SC_example_reflect}

In the Majorana representation the root state Hamiltonian
of a three-dimensional class BDI superconductor with reflection symmetry is given by
 \begin{eqnarray} \label{examp_3_BDI_SC_embed}
&&\mathcal H^{(0)}=  i\partial_x X_{303}+i\partial_y X_{103}+i\partial_z X_{001}+m X_{002}, 
\end{eqnarray}
where $X_{ijk}$ denotes the tensor product of three Pauli/identity matrices. 
This Hamiltonian is invariant under time-reversal symmetry and reflection symmetry $x \to -x$ with
the symmetry operators
\begin{eqnarray}
  \mathcal T=X_{223}  \mathcal K \quad \textrm{and}  \quad R_x=X_{100},  
\end{eqnarray}
respectively. We note that in the Majorana representation PHS with operator $C$ is implemented trivially. 
(Here, TRS with $\mathcal{T}^2=+1$ could be viewed as a combination of a $\pi$ spin-rotation symmetry times a TRS $\mathcal{\hat T}$ with
$\mathcal{\hat T}^2=-1$ for spin-1/2 particles.)
Since $\mathcal T^2= +1$, $C^2 =+1$, $[R, \mathcal T ] = -1$, and $[R,  C ] = +1$, 
Hamiltonian \eqref{examp_3_BDI_SC_embed} belongs to class BDI with $R_{-+}$ in Table~\ref{tab:results}.


The rank of the root state Hamiltonian~\eqref{examp_3_BDI_SC_embed} is $r_{\textrm{min}}=8$. Since the boundary Hamiltonian of the superconductor~\eqref{examp_3_BDI_SC_embed} has no continuous symmetry, its associated parent symmetry class is class D.
The two-dimensional root state Hamiltonian of parent symmetry class D has rank $r_m=2$, because
the associated Clifford algebra is $Cl_{1,2}$, whose matrix representation is $2 \mathbb{R} (2)$ with rank two.
From Table \ref{tab:minimal copies}, we infer that in $d-1=2$ spatial dimensions 
 $\nu_m = 2^4=16$ copies of the class~D root state can be continuously connected to the trivial state.
 Hence,  according to Eq.~\eqref{main_reduction_formula}, the classification of Hamiltonian~\eqref{examp_3_BDI_SC_embed} is
 $\mathbb Z_{\frac{2\nu_mr_m}{r_{min}}} = \mathbb Z_8$.
 That is, for eight copies of the root state Hamiltonian~\eqref{examp_3_BDI_SC_embed} the surface states can be gapped
 out by quartic interactions, which is in agreement with Table~\ref{tab:results}.

Let us now explicitly construct the allowed Dirac masses for the boundary Hamiltonian of Eq.~\eqref{examp_3_BDI_SC_embed}.
The boundary Hamiltonian is derived by considering a domain wall along the $z$-direction in the mass term~$m X_{002}$.
Introducing quartic interactions and performing a Hubbard-Stratonovich transformation yields
\begin{eqnarray}
\label{eq:1}
&&H_{bd}^{(dyn)}=(i\partial_x X_{30}+i\partial_y X_{10})\otimes \mathds 1+M(\tau,x,y) ,
\end{eqnarray}
where the mass term $M ( \tau , x, y )$ is a $4\nu\times4\nu$ matrix,
with $\nu$ the number of root state copies.
On the boundary, the operators for TRS and
reflection symmetry are given by
\begin{eqnarray} \label{symmetries_for_bd_ham_BDI_SC_embed}
&&\mathcal{T}_{bd} =X_{22}\mathcal K \quad \textrm{and} \quad R_{bd, x} =X_{10},
\end{eqnarray}
respectively.
Generic symmetry-preserving contact interactions lead to a dynamical boundary mass term $M(\tau,x,y)$  in symmetry class D.
Therefore, we can parametrize the space of the dynamical mass matrices as $M (\tau, x, y) =\sigma_2 \otimes \tilde M(\tau, x,y)$, where  $\tilde M$
is a $2 \nu \times 2 \nu$ real-valued and symmetric matrix.
The space of the matrices $\tilde M$
 is topologically equivalent to 
\begin{equation}
V_\nu=\cup_{n=0}^{2\nu}O(2\nu)/[O(2\nu-n)\times O(n)], 
\end{equation}
which in the limit $\nu\rightarrow \infty$ becomes the classifying space $R_0$. 
Similar to the previous two examples, we now explicitly construct 
the allowed dynamical boundary masses for the copy numbers $\nu=2^n$, with $n=0, 1,2, 3$, in the following.

\emph{Case $\nu=1$.} ---
For $\nu=1$, the space of the mass matrices $M ( \tau, x, y)$ is spanned by the pair of anti-commuting matrices
$X_{21}$ and $X_{23}$. (There does not exist a third mass term since $\pi_0(R_1)=\mathbb Z_2$.)
Thus, the QNLSM target space is $S^{N(1)-1} = S^1$. Because $\pi_1 ( S_1) = \mathbb{Z}$, there exists a topological
obstruction of the vortex type, which prevents the gapping of the surface states.

\emph{Case $\nu=2$.} ---
For $\nu=2$ there exist only $N(2) = 3$ pairwise anti-commuting mass matrices, since  $\pi_0(R_2)=\mathbb Z_2$, namely
 $X_{213}$, $X_{233}$, and $X_{201}$.
 The space spanned by these three mass matrices is homeomorphic to the two-sphere $S^2$.
 Since $\pi_2 (S^2) = \mathbb{Z}$, $M ( \tau, x, y)$ can support monopole defects. That
 is the QNLSM possesses a topological term of the monopole type and, hence,
 the surface modes cannot be gapped out.
  
\emph{Case $\nu=4$.} ---
For four copies  $\nu=4$, we find the five pairwise anti-commuting Dirac masses $X_{2333}$, $X_{2331}$, $X_{2130}$, $X_{2122}$,
and $X_{2010}$. (There does not exist a sixth Dirac mass since $\pi_0(R_4)=\mathbb Z$.)
These five matrices span the space of the mass matrices $M ( \tau, x, y)$, which is homeomorphic
to the four-sphere $S^4$. That is, the QNLSM target space is given by $S^{N(4)-1} = S^4$
Because $\pi_4 (S^4) = \mathbb{Z}$, it is possible to add a WZ topological term to the QNLSM and, hence,
the surface states remain gapless in in the presence of interactions.
 
\emph{Case $\nu=8$.} ---
For $\nu=8$ one finds that there exist nine pairwise anti-commuting Dirac masses.  (This is becuase the next nontrivial homotopy group is $\pi_0(R_8)=\mathbb Z$.) Hence, the QNLSM target space is $S^{N(8)-1} = S^8$. Since $\pi_\iota ( S^8) = 0$, for $\iota = 0, 1, \ldots, 4$, it is not possible to add a topological term to the QNLSM. As a consequence the surface modes can be gapped out by interactions.

Therefore, the classification of Hamiltonian~\eqref{examp_3_BDI_SC_embed} reduces from $\mathbb{Z}$ to $\mathbb{Z}_8$, in agreement with the derivation given above.

\subsubsection{BDI insulator with reflection symmetry}
\label{BDI_TI_example_reflect}

Let us now interpret the class BDI topological state as an insulator with U(1) charge conservation, i.e., as a topological insulator 
 with particle-hole symmetry, time-reversal symmetry that squares to $+1$, and U(1) symmetry. In other words, the protecting symmetries are 
 U(1)$\rtimes [Z_2^C\times \mathcal T(1)]$. 
In order to implement these symmetries the rank of the root state~\eqref{examp_3_BDI_SC_embed} needs to be doubled.
We obtain 
\begin{eqnarray} \label{examp_3_BDI_TI_embed}
\label{eq:large}
\mathcal H^{(0)}=i\partial_x X_{3010}+i\partial_y X_{1010}+   i\partial_z X_{0022}+m X_{0032}  ,
\end{eqnarray}
with the symmetry operators
\begin{eqnarray}
 \mathcal T=X_{2203} \mathcal K,  \; R_x=X_{1000}, \; C=X_{0013}, \;  \textrm{and} \;   Q=iX_{0002}, \qquad
\end{eqnarray}
where $Q$ is the generator of the continuous U(1) symmetry. 

The rank of the root state~\eqref{examp_3_BDI_TI_embed} is $r_{\textrm{min}}=16$. Since the boundary Hamiltonian
of Eq.~\eqref{examp_3_BDI_TI_embed} exhibits a U(1) continuous symmetry, the parent symmetry class that we need to consider 
is class A. (In this case, the space of the dynamical boundary masses is topologically equivalent to $\cup_{n=0}^{2\nu}U(2\nu)/[U(2\nu-n)\times U(n)]$, 
which in the limit $\nu\rightarrow\infty$  corresponds to the classifying space $C_0$.)
The rank of the two-dimensional root state of parent symmetry class A is $r_m=4$, since 
 $\dim Cl_{2}=4$.
From Table~\ref{tab:minimal copies} we find that $\nu_m= 2^3=8$ copies of the two-dimensional class A root state are needed to
gap out the edge modes. Hence, if we allow only for U(1) symmetric dynamical masse, then the reflection-symmetric BDI topological insulator~\eqref{examp_3_BDI_TI_embed} has 
a $\mathbb Z_{\frac{2\nu_mr_m}{r_{min}}} = \mathbb Z_4$ classification. 

Upon relaxing the constraints from the U(1) symmetry, the dynamical masses fall into class D.
(In this case the space of the dynamical masses is equivalent to $\cup_{n=0}^{4\nu}O(4\nu)/[O(4\nu-n)\times O(n)]$,
which in the limit $\nu\rightarrow\infty$  becomes the classifying space $R_0$.) 
The rank of the two-dimensional root state in class D is $r_m=2$. By use of Table~\ref{tab:minimal copies}, one finds
that $\nu_m = 2^4=16$ copies of the root state can be connected to the trivial state.
Hence, the classification is again $\mathbb Z_{\frac{2\nu_mr_m}{r_{min}}} = \mathbb Z_4$
(even without checking the invariance of the target space under U(1) operation). 
With this we conclude that the reflection-symmetric BDI topological insulator~\eqref{examp_3_BDI_TI_embed} is indeed 
classified as $\mathbb Z_4$  (cf.~caption of Table~\ref{tab:results}.)
This is in contrast to the reflection-symmetric BDI topological superconductor~\eqref{examp_3_BDI_SC_embed}
which is classified as $\mathbb Z_8$.


\subsubsection{Bosonization analysis for the boundary Hamiltonian}
\label{examp_BDI_bosnization_analysis}

In this section we use the bosonization technique to perform a stability analysis of the 
surface states of the BDI superconductor~\eqref{examp_3_BDI_SC_embed} and the BDI insulator~\eqref{examp_3_BDI_TI_embed}.
We will see that the classification obtained from this stability analysis agrees with the QNLSM appraoch.

\paragraph{BDI superconductor with reflection symmetry}
We first consider the BDI superconductor~\eqref{examp_3_BDI_SC_embed}.
Following  Refs.~\onlinecite{isobe2015, yoshida2015}, we introduce a spatial modulation in the
mass term of the boundary Hamiltonian~\eqref {eq:1}. That is, we consider the boundary Hamiltonian
\begin{eqnarray} \label{bd_ham_for_bosonization}
&&H_{bd}=i\partial_x X_{30}+i\partial_y X_{10} +m(x ) X_{23} ,
\end{eqnarray}
where the mass term $m(x)=m_0 sgn(x)$ describes a domain wall with a kink at $x=0$.
Observe that $H_{bd}$, Eq.~\eqref{bd_ham_for_bosonization}, satisfies both TRS and reflection symmetry $x \to -x$ with
the symmetry operators given by Eq.~\eqref{symmetries_for_bd_ham_BDI_SC_embed}.
(In passing we note that the surface Hamiltonian~\eqref{bd_ham_for_bosonization} with a spatially independent mass term $m \equiv m_0$
can be viewed as a two-dimensional bulk system with TRS and an internal $Z_2$ symmetry with operator $X_{03}$.
In fact, there exists a general connection between $d$-dimensional systems with reflection symmetry 
and $(d-1)$-dimensional systems with an internal $Z_2$ symmetry, see Appendix~\ref{appendixC} for more details). 
In the presence of the domain wall $m(x)$, the surface  Hamiltonian~\eqref{bd_ham_for_bosonization} exhibits two counter-propagating helical modes
that are localized at the kink of the domain wall $x=0$.
The dynamics of the these two  gapless modes  is described by the low-energy Hamiltonian
\begin{eqnarray}
\label{eq:domain}
&&H_{dw}=i\partial_y X_3 .
\end{eqnarray}
The two helical modes at the domain-wall  transform into each other under TRS (with operator $\mathcal T=X_{1}\mathcal K$) and
are symmetric under reflection $x \to -x$ with operator $R_x=X_3$.

We now use bosonization to study the stability of the gapless domain-wall states in the presence of interactions. 
Taking two copies of the system, we combine two gapless Majorana domain-wall modes with a given propagation direction to form one complex fermion mode.
These complex fermion modes are then converted into  bosonic fields ${\boldsymbol \phi}  = (\phi_1,\phi_2)^T$ using the standard bosnization procedure~\cite{wen_IJMPB_92,wen_zee_PRB_92}.
The Lagrangian for these bosonic fields describing the domain-wall modes is given by
\begin{eqnarray} \label{lagrangian_bosonic_fields}
\mathcal L=\int \frac{dx}{4\pi} [K_{I,J} \partial_t\phi_{I}(x)\partial_x\phi_J(x)-\partial_x\phi_I(x)\partial_x\phi_I(x)] , \quad
\end{eqnarray}
where $K$ is the third Pauli matrix and summation over repeated indices is assumed.
The bosonic fields ${\boldsymbol \phi}  = (\phi_1,\phi_2)^T$ represent domain-wall modes moving in the $+y$ and $-y$ directions, respectively. 
That is, the vertex operators $:e^{i\phi_1}:$ and $:e^{i\phi_2}:$ create left- and right-moving fermionic modes. (Here, the colons
denote a normal-ordered operator, as usual.)
The commutation relations among the bosonic fields are given by
\begin{eqnarray}
[\phi_I(x),\phi_J(y)]=i\pi K_{I,J} \, \textrm{sgn} (x-y)+i\pi  \,\textrm{sgn} (I-J) .
\end{eqnarray}
From Eq.~\eqref{eq:domain}, we infer that TRS and rotation symmetry act on the bosonic fields as
\begin{subequations} \label{eq:sym}
\begin{eqnarray}
\mathcal T  {\boldsymbol \phi}(x) \mathcal T^{-1} &=& -\sigma_1 {\boldsymbol \phi} (x),  \\
R_x {\boldsymbol \phi} (x) R_x^{-1} &=&  {\boldsymbol \phi} (-x)+\pi \mathbf e_2 ,
\end{eqnarray}
\end{subequations}
where $\mathbf e_i$ denotes the unit vector whose $i$th entry is one and the other entries are zero. 

Let us now examine whether interactions can gap out $\nu$ copies of the gapless helical domain-wall modes, described by Lagrangian~\eqref{lagrangian_bosonic_fields}, without breaking the symmetries. (Observe that $\nu$ copies of Lagrangian~\eqref{lagrangian_bosonic_fields}  correspond
to $2\nu$ copies of the original system, Eq.~\eqref{bd_ham_for_bosonization}.)
Interactions among the domain-wall modes, such as backscattering and umklapp processes, are described 
by cosine terms of the form  
\begin{eqnarray} \label{lagarangian_bosoniz_int}
\mathcal L_{int}= \sum_{\alpha=1}^\nu C_\alpha \int dx :\cos \left( \mathbf l_{\alpha}\cdot {\boldsymbol \phi}+a_{\alpha} \right): ,
\end{eqnarray} 
where $C_{\alpha}$ and $a_{\alpha}$ denote  real-valued coupling constants and  phase factors, respectively. 
The vectors $\mathbf l_{\alpha}$ ($\alpha=1, \ldots, \nu$) are a set of $v$ independent integer-valued vectors, chosen
such that  $\mathcal L_{int}$ respects all symmetries and
the fields satisfy~\cite{haldane_PRL_95} $[ \mathbf l_{\alpha} \cdot {\boldsymbol \phi (x) } ,  \mathbf l_{\beta} \cdot {\boldsymbol \phi} (y) ]=0$ up to
$2 \pi i n$, with $n \in \mathbb{Z}$.
Furthermore, to ensure that there is no spontaneous symmetry breaking, the set of elementary bosonic variables\cite{lu2012,yoshida2015,levin_stern_fractional_TI}
$\{ {\bf v}_\alpha \cdot \boldsymbol{\phi} \}$ must stay invariant modulo $2\pi$ under the symmetry transformations in Eq.~\eqref{eq:sym}.
With these conditions, we find that for $\nu=4$ copies of  $\mathcal L$ the  domain-wall states can be gapped out
by the symmetry preserving interactions~\eqref{lagarangian_bosoniz_int} with the gapping vectors~$\mathbf l_\alpha$ given by
\begin{eqnarray} \label{gapping_vectors_A}
\mathbf l_1 &=& (1,0 \, | \, 1,0 \, | \, 0,-1 \, | \,  0,-1)^T, \nonumber\\
\mathbf l_2 &=& (0,1 \, | \, 0,1 \, | \, -1,0 \, | \, -1,0)^T, \nonumber\\
\mathbf l_3 &=& (1,-1 \, | \, -1,1 \, | \, 0,0 \, | \, 0,0)^T, \nonumber\\
\mathbf l_4 &=& (0,0 \, | \, 0,0 \, | \, 1,-1 \, | \, -1,1)^T, 
\end{eqnarray}
and with all $a_\alpha$'s equal to zero and $C_\alpha = 1$.
In Eq.~\eqref{gapping_vectors_A}, the vertical lines separate copies of helical edge modes. 
It is easy to check that the gapping vectors~\eqref{gapping_vectors_A} satisfy the symmetry constraints and all other necessary conditions.
Hence, for $2 \nu=8$ copies of the BDI superconductor~\eqref{examp_3_BDI_SC_embed} [i.e., $\nu =4$ copies of  $\mathcal L$, Eq.~\eqref{lagrangian_bosonic_fields}]  the surface modes are completely gapped out by the interaction \eqref{lagarangian_bosoniz_int} with \eqref{gapping_vectors_A}.
Therefore, three-dimensional BDI superconductors with reflection symmetry form a $\mathbb Z_8$ group, which is in agreement with the QNLSM approach of Sec.~\ref{BDI_SC_example_reflect}.


\paragraph{BDI insulator with reflection symmetry}
A similar analysis can be performed for the BDI insulator~\eqref{examp_3_BDI_TI_embed},
in which case the defining symmetries are U(1)$\rtimes [Z_2^C\times \mathcal T]$. 
To this end, we first rewrite Hamiltonian~\eqref{eq:large} in complex fermion representation, in which the rank 
of the Hamiltonian is halved. We find 
\begin{eqnarray}
&&\mathcal H^{(0)}=i\partial_x X_{301}+i\partial_y X_{101} +i\partial_z X_{002}+m X_{003}  .
\end{eqnarray}
Within the complex fermion representation the U(1) charge conservation symmetry with generator $Q$ is realized trivially.
The operators of TRS, reflection, and PHS are given by
\begin{eqnarray}
&&\mathcal T=X_{220}\mathcal K , \quad R_x=X_{100}, \quad \textrm{and} \quad C=X_{001}\mathcal K ,
\end{eqnarray}
respectively. 
Following similar steps as above, we first introduce a domain wall along the $z$-direction in the mass term $m X_{003}$
to derive the surface Hamiltonian.
Subsequently, we consider an odd-parity spatial modulation in the mass term $ m  X_{23}$ of the surface Hamiltonian, i.e.,  
$  m_0 \textrm{sgn} (x) X_{23} $.
In the presence of the domain wall $m_0 \textrm{sgn} (x) $ the surface Hamiltonian exhibits two counter-propagating helical modes localized at 
the kink of the domain wall $x=0$. The low-energy dynamics of these two helical modes is described by Hamiltonian~\eqref{eq:domain}, except that  
now we are using the complex fermion representation.
 
 Using the bosonization procedure, the two counter-propagating complex modes at the domain wall are transformed into two bosonic fields denoted by
 ${\boldsymbol \phi}  = (\phi_1,\phi_2)^T$. Under TRS and reflection symmetry the bosonic fields transform according to Eq.~\eqref{eq:sym}, just as before. In the present case, there are two additional constraints due to U(1) charge conservation and PHS, which are implemented by
\begin{subequations}
 \begin{eqnarray}
e^{i\tilde Q\theta} \boldsymbol{ \phi } e^{-i\tilde Q\theta} &=& \boldsymbol{\phi} +\theta (\mathbf e_1+ \mathbf e_2) ,\\
C \boldsymbol{ \phi } C^{-1} &=& - \boldsymbol{ \phi} ,
\end{eqnarray}
\end{subequations}
where $\tilde Q$ denotes the generator of  the U(1) symmetry written in the complex fermion representation. 
As it turns out, for $\nu =4$ copies of the surface domain-wall Hamiltonian the helical edge modes can be
gapped out by  interaction $\mathcal{L}_{int}$, Eq.~\eqref{lagarangian_bosoniz_int}, with the same gapping vectors~\eqref{gapping_vectors_A} as above.
One can check that the gapping vectors~\eqref{gapping_vectors_A} satisfy  all symmetry constraints.
Hence, the classification of  three-dimensional BDI insulators with reflection symmetry reduces  to 
$\mathbb{Z}_4$, in agreement with Sec.~\ref{BDI_TI_example_reflect}.

\section{Conclusions}
\label{sec_conclusion}
In this paper, we  have determined, in all generality,
whether the surface states of topological crystalline insulators (TCIs) and topological crystalline superconductors (TCSCs)   
with order-two symmetries  (i.e., reflection or twofold rotation) are stable in the presence of quartic fermion-fermion interactions.
To achieve this, we have described the interaction effects on the surface states in terms of a quantum non-linear $\sigma$ model (QNLSM),
whose target space is derived from  Clifford algebra extensions (see Sec.~\ref{sec_collapse_reflection}).
Whether the boundary modes can be gapped out by symmetry-preserving interactions depends on the presence or absence of a topological obstruction (i.e., a topological term) in the action of the QNLSM. 
The existence of this topological term, in turn, follows from the 
 homotopy group of the QNLSM target space.
By performing this  analysis for multiple copies of a given topological  phase,
we have derived a systematic  classification of interacting topological crystalline insulators and superconductors, which is summarized in Tables~\ref{tab:results} and~\ref{tab:results2}. 
Interestingly, the noninteracting $\mathbb{Z}_2$ classifications are
stable in the presence of interactions, while the $\mathbb{Z}$ classifications  
reduce  to $\mathbb{Z}_N$, see Eq.~\eqref{main_reduction_formula}.
 
Tables~\ref{tab:results} and \ref{tab:results2} contain many interesting TCIs/TCSCs with a reduced classification in physical dimensions
$d=1$, $d=2$, and~$d=3$. For three of these we have discussed explicit examples in Sec.~\ref{section_examples},
namely, a Majorana wire with two-fold rotation symmetry, a two-dimensional reflection-symmetric spin-singlet superconductor,
and a three-dimensional BDI insulator/superconductor with reflection symmetry.
Some of the entries in Table~\ref{tab:results} describe TCIs/TCSCs that have been previously studied in the literature,
e.g., the  two-dimensional DIII superconductor with reflection symmetry~\cite{yao2013,morimoto2015} [DIII+$R_{--}$, reduced to $\mathbb Z_8$] and the three-dimensional AII insulator with reflection symmetry~\cite{Hsieh:2012fk,morimoto2015} [AII+$R_-$, reduced to $\mathbb Z_{8}$], which
is physically realized in the rocksalt SnTe~\cite{Tanaka:2012fk,Hsieh:2012fk,Xu:2012} and in the antiperovskites A$_3$PbO~\cite{Hsieh_antiperovskite_12,chiu_antiperovskite_2016}.
It would be exciting to experimentally verify the interaction-induced collapse of the free-fermion classification in a physical system.
 Particularly suited for this purpose are one-dimensional systems, e.g., the Majorana chain with two-fold rotation symmetry
discussed in Sec.~\ref{kitaev_majorana_chain_with_rotation}.
This TCSC could be realized, for example, in terms of
Shiba bound states induced by magnetic adatoms on the surface of an $s$-wave superconductor~\cite{NadjPerge602}.
Another suitable system to experimentally test the classification collapse is 
the Su-Schrieffer-Heeger (SSH) dimer chain with two-fold rotation symmetry~\cite{su1979}, which belongs
to class BDI with $U_{++}$ in Table~\ref{tab:results2}\footnote{It collapses to $\mathbb Z_4$ because there is an additional U(1)$\rtimes Z_2^C$ symmetry here.}. 
It has recently become possible to fabricate the SSH dimer chain in designer platforms, for example, using
cold atoms~\cite{meier_gadway_Ncomms_16} or chlorine vacancy lattices on top of Cu(100)~\cite{drost_ojanen_arXiv_16}.
Further progress in this direction may allow to fabricate multiple SSH chains and to study the interactions among them.



\acknowledgments
We thank Yi-Zhuang You and Y.~X.~Zhao for helpful discussions.
 X.-Y.S. benefitted from the lectures and discussions during the 2016 Boulder Summer School for Condensed Matter and Materials Physics.
The support by the KITP at UC Santa Barbara is gratefully acknowledged. This research was supported in part by the National Science Foundation under Grant No.~NSF PHY-1125915.

\appendix
\section{Symmetries of many-body Hamiltonian \& connection between real Majorana and
complex fermion representations}
\label{appendix_A}

In this appendix we show how the symmetries act on the many-body Hamiltonian written in terms of complex
fermion operators or real Majorana operators (Sec.~\ref{symmetry}). We also 
show that the reduction patterns of the free-fermion classifications can be derived
using both the real Majorana and the complex fermion representations.
Both representations give consistent results  (Sec.~\ref{connection}).

\subsection{Symmetries of many-body Hamiltonian}
\label{symmetry}

In the complex fermion basis, we write a generic gapped fermionic many-body Hamiltonian
\begin{equation}
\mathcal H=\int d^d \mathbf x\int d^d \mathbf y \sum_{ij} \Psi_i^\dagger(t,\mathbf x) H_{ij}(\mathbf x,\mathbf y) \Psi_j(t,\mathbf y)
\end{equation}
where the second quantized fermionic operators obey the canonical equal time anticommutation relations.

Time-reversal symmetry (TRS) $\mathcal T=T\mathcal K$ ($\mathcal K$ denotes complex conjugation) acts on the operator level as
\begin{equation}
\mathcal T \psi_j(t,\mathbf y) \mathcal T^{-1}=T^*_{j'j}\psi_{j'}(-t,\mathbf y)
\end{equation}

If we assume translation invariance in the system and consider mapping the real-space Hamiltonian into reciprocal space, TRS requires
\begin{equation}
TH^*(\mathbf k)T^{-1}=H(-\mathbf k)
\end{equation}
which in the notation of Eq.~\eqref{eq:hamiltonian} amounts to $\{\tilde\gamma_i,\mathcal T\}=0,[\tilde\beta, \mathcal T]=0$ if we consider the massive Dirac Hamiltonian. In addition, $T^*T=\pm 1$ distinguish two different TRS.

Particle-hole symmetry (PHS, also called charge-conjugation symmetry) is a unitary symmetry which reverses the sign of the fermion number $\psi^\dagger_i(\mathbf x)\psi_i(\mathbf x)-\frac{1}{2} \delta (\mathbf x=0)$ and acts on the operator level as
\begin{equation}
C\psi_j(t,\mathbf y)C^{-1}=C_{j'j}\psi^\dagger_{j'}(t,\mathbf y)
\end{equation}

Assuming the Hamiltonian is traceless, one could verify that PHS requires 
\begin{equation}
CH^*C^{-1}=-H
\end{equation}
namely PHS is realized \emph{anti-unitarily} on the first-quantized Hamiltonian. One could formally write PHS as $\mathcal C=C\mathcal K$ to represent its operation on the first-quantized Hamiltonian. PHS dictates that $\{\tilde\beta,C\}=0,[\tilde\gamma_i,C]=0$ for the Dirac Hamiltonian. $C^*C=\pm 1$ distinguish two different PHS.

Chiral symmetry (CHS) is an anti-unitary symmetry $\mathcal S=\Gamma\mathcal K$ that combines TRS and PHS. It's realized as
\begin{equation}
\mathcal S \psi_j(t,\mathbf y) \mathcal S^{-1}=\Gamma_{j'j}\psi_{j'}^\dagger(t,\mathbf y)
\end{equation}

Assuming traceless condition of the Hamiltonian, CHS dictates the condition on the first-quantized Hamiltonian $\Gamma H\Gamma^{-1}=-H$, which means $\{\Gamma,\tilde\gamma_i\}=\{\Gamma,\tilde\beta\}=0$. We note that it's unitarily realized in the first-quantized Hamiltonian level.

When writing a BdG Hamiltonian, we arrange the $\Psi$ as Nambu spinors $\Psi=(\psi_1,\psi_2,\cdots \psi_N,\psi_1^\dagger,\psi_2^\dagger,\cdots \psi_N^\dagger)^T$. This renders $\Psi$ and $\Psi^\dagger$ as not independent from each other, $\Psi=\sigma_1 (\Psi^\dagger)^T$ ($\sigma_1$ acts in the Nambu space), which is in the form of PHS. So the BdG Hamiltonian has a ``built-in" particle-hole symmetry $\sigma_1 H^*\sigma_1=-H$. This symmetry is actually trivially realized written in the Majorana basis.

Working in the real Majorana basis $\{\chi_a\}$, where the fermion annihilation operator is written as $\psi_i=\chi_{2i-1}+i\chi_{2i}$, we write down a generic Dirac Hamiltonian in $d$ spatial dimension
\begin{equation}
\mathcal H=i\chi_a [\sum_{i=1}^d(\partial_i\tilde\gamma_i)_{ab}+m \tilde\beta_{ab}]\chi_b
\end{equation}
with real symmetric kinetic matrices $\{\tilde\gamma_i\}$ satisfying $\{\tilde\gamma_i,\tilde\gamma_j\}=2\delta_{ij}$ and they all anticommute with real anti-symmetric mass matrix $\tilde\beta$. We could flatten the spectrum by choosing $(m\tilde\beta)^2=-1$.

A global U(1) symmetry takes $\chi\rightarrow e^{Q\theta}\chi$ where $Q$ is a real anti-symmetric matrix satisfying $[Q,\tilde \gamma_i]=[Q,\tilde\beta]=0, Q^2=-1$.

A unitary $Z_2$ symmetry $\mathbf C$ is represented by a real matrix $C$ satisfying $[C,\tilde\gamma_i]=[C,\tilde\beta]=0, C^TC=1, \{Q,C\}=0$. If $Q$ corresponds to charge conservation and $\mathbf C$ corresponds to particle-hole symmetry, we have $C^2=1$; on the other hand, if $Q$ represents $S^z$ spin conservation and $\mathbf C$ is the generator for $S^y$ spin rotation [$\chi\rightarrow e^{C\theta} \chi$], then $C^2=-1$.

TRS is written as $\mathcal T=T\mathcal K$ with a real matrix $T$ satisfying $\{T,\tilde\gamma_i\}=\{T,\tilde\beta\}=0, T^TT=1$. $T^2=\pm 1$ depends on whether $T$ is symmetric or anti-symmetric. $T$ may commute/anti-commute with $Q$ depending on the specific symmetry group.

The PHS could be either a real $Z_2$ particle-hole symmetry with $C^2=\pm 1$ or a fictitious one representing a continuous spin rotation symmetry $\chi\rightarrow e^{C\theta}\chi$ satisfying $C^2=-1$  with the above U(1) symmetry identified to be the spin rotation symmetry around another axis [these together enforce the $SU(2)$ symmetry of the system, with the third generator of spin rotation being $QC$]. We further have $\{Q,T\}=0$ when $Q$ corresponds to particle number; while in the case of $SU(2)$ spin rotation symmetry, thing are more complicated: In the case of $SU(2)$ spin rotation, when $\mathcal T$ is physical TRS, we have $[Q,T]=[C,T]=0$; when $\mathcal T$ is the combination of TRS and $\pi$ spin rotation, we could always choose to make $[C,T]=0$ while dictating $\{Q,T\}=0$; this corresponds to the second embedding scheme of class $CI$ denoted as U(1)$\rtimes [Z_2^C\times T]$ in the explanation column of Table \ref{tab:symmetries}.
It's also verified that we can always choose to have $[T,C]=0$. 

The reflection symmetry is represented as, say, $R_xP$ where $P$ represents the operation in real space that takes $x\rightarrow -x$ and $R_x$ is the matrix acting on internal degrees of freedom. It requires $\{R_x, \tilde\gamma_x\}=0, [R_x,\tilde\gamma_i(i\neq x)]=0,[R_x,\tilde\beta]=0$.

For two-fold rotation symmetry $U$, the invariance of the Hamiltonian Eq.\eqref{eq:hamiltonian} under this rotation symmetry 
\begin{equation}
U^{-1}\mathcal H^{(0)}(k_1,k_2,\cdots, k_d)U= \mathcal H^{(0)}(-k_1,-k_2,\cdots, k_d)
\end{equation}
 dictates that $\{U,\tilde\gamma_i(i\neq d)\}=0,[U,\tilde\gamma_d]=[U,\tilde\beta]=0$.
 
\subsection{Connection between real Majorana and complex fermion representations}
\label{connection}

While the symmetry conditions for ``AZ" symmetry classes in terms of complex fermions are long well-known , there's ambiguity concerning whether there's additional U(1) symmetry [depending on whether it's written in terms of Nambu spinor form] and whether the PHS is real $Z_2$ particle-hole symmetry or a fictitious one coming from, say, continuous spin rotation symmetry \cite{chiu2015}. While in the Majorana basis, we could resolve the uncertainties.

The U(1) symmetry corresponds to a nontrivial orthogonal transformation in Majorana basis $\chi\rightarrow e^{Q\theta}\chi (\theta\in [0,2\pi))$ with $Q$ being a real anti-symmetric matrix with $Q^TQ=1$. There's a conserved ``particle number" $N=i\chi_bQ_{ba}\chi_a$ [repeated indices are assumed to be summed over]. The eigenvectors of $Q$ corresponding to eigenvalues $\pm i$ are $\eta_{\pm j}$'s satisfying $[N, \eta_{\pm j,a}\chi_a]=\pm \eta_{\pm j,a}\chi_a$, which have one-to-one correspondence $i$ and $-i$ eigenvalues by complex conjugation of their coefficients. The conserved quantity dictates that the Hamiltonian only involves fermion bilinears in form of $(\eta_{+ j,a}\chi_a)(\eta_{- k,b}\chi_b)$. Reorganizing the Majoranas into complex fermions $\psi_i=\eta_{+i,a}\chi_a (\psi_i^\dagger=\eta_{-i,a}\chi_a)$, the first quantized Hamiltonian could be written as $\Psi^\dagger H^{(0)} \Psi$ with \emph{the dimension of $H^{(0)}$ in complex fermion basis halved by virtue of U(1) symmetry}.

Another noteworthy point is that since $\{Q,C\}=0$, the operation of $C$ will take an eigenvector $\eta_{\pm i,a}\chi_a$ of $Q$ to $\eta_{\mp j,a}\chi_a$ and loosely amounts to $\Psi\rightarrow ( \Psi^\dagger)^T$, which is consistent with the definition of PHS in complex basis. If reflection also anticommutes with $Q$, then in the same logic it's also anti-unitarily realized in the first-quantized level. We note that $\gamma_xR_x$ (assuming reflection in the $x$ direction) is equivalent to a \emph{global} TRS in this case. So if TRS is absent in the original symmetry class, the scenario will become another AZ class with an additional TRS ($\mathcal T^2$ depends on $R_x^2$), which has already been resolved in previous work; if TRS is present, then two anti-unitary symmetries is equivalent up to a global unitary symmetry which can be made trivial by block diagonalizing Hamiltonian in terms of the irreducible representation spaces.\cite{ludwig2016,chiu2015} So we only consider cases where $[Q,R]=0$.

If one wants to work in complex fermion basis to deduce the reduction pattern, in principle one can follow the same procedure outlined in section \ref{procedure} with several caveats. The rank of root state in complex basis should be determined by virtue of its relation to that written in Majorana basis stated above. For cases with U(1) or SU(2) continuous symmetries, it should be kept in mind to include ``superconducting fluctuations" by a particle-hole grading discussed in section \ref{two basis}. We check by explicit calculation that complex basis yields the same results as that in Majorana basis.

\section{Relevant Clifford algebra for the 27 cases}
\label{clifford}


In this section, we briefly overview how to represent the kinetic/mass matrices along with symmetry operations as the generators of Clifford algebras\cite{morimoto2013,lu2014} and therefore determine the rank of their matrix representation (hence the size of the root states).

We first consider writing in complex fermion basis. Introducing an ``imaginary unit" $J$ that anticommutes with TRS  and PHS with $J^2=-1$. The original Clifford algebra for the ten symmetry classes without reflection symmetry is as follows (we take $\gamma_i,M$ to represent $\tilde\gamma_i,\tilde\beta$ below. TRS and PHS can be made to commute with each other.):

i) For complex class A: $\{\gamma_i,M\}$ constitutes a complex Clifford algebra $Cl_{d+1}$. For class AIII, $\{\gamma_i,M,\Gamma\}$ constitutes a complex clifford algebra $Cl_{d+2}$.

ii) For classes with only TRS (AI,AII): $\{\gamma_i,JM,T,TJ\}$ constitutes a real Clifford algebra $Cl_{1,d+2}(AI), Cl_{3,d}(AII)$.

iii) For classes with only PHS (C,D): $\{J\gamma_i, M,C,CJ\}$ constitutes a real Clifford algebra $Cl_{2+d,1}(C),Cl_{d,3}(D)$.

iv) For classes with both symmetries (BDI,DIII,CI,CII): $\{J\gamma_i, M,C,CJ,TCJ\}$ constitutes a real Clifford algebra $Cl_{d+1,3}$(BDI),$Cl_{d,4}$(DIII),$Cl_{d+2,2}$(CI),$Cl_{d+3,1}$(CII).

With reflection symmetry $R_x$, we note that $i\gamma_x R_x$ anticommutes with all other matrices in the Hamiltonian. In class AIII, if reflection anticommutes with CHS, then $\Gamma \gamma_x R_x$ commutes with all the generators in the original Clifford algebra, which won't enlarge the Clifford algebra; else $\gamma_x R_x$ is a new generator. For the other cases with no or only one protecting anti-unitary symmetry, we could always add $\gamma_x R_x$ or $J\gamma_x R_x$ to the original complex/real clifford algebra to form the new Clifford algebra [note that $J$ will change the anti/commutation relation to TRS/PHS, so we could always manage to make this new element anticommute with the generators containing symmetry operators]. For the cases with both TRS and PHS, if reflection symmetry anti/commutes with \emph{both} the two symmetries, one can verify that either $\gamma_x R_x$ or $J\gamma_x R_x$ could serve as a new generator. In the case of $R_{-+},R_{+-}$, either the generator $\tilde M=TC\gamma_x R_x$ or the generator $\tilde M=JTC\gamma_x R_x$ commutes with all the original generators. If $\tilde M^2=1$, then this won't change the original relevant Clifford algebra. If $\tilde M^2=-1$, this would change the original real Clifford algebra $Cl_{p,q}$ to a complex one $Cl_{p+q}$ [$Cl_{p,q}\otimes Cl_{1,0}\simeq Cl_{p+q}$]. The complete Clifford algebra is listed at the first in the third column of Table \ref{tab:results}.

Next we state how to incorporate reflection symmetry in Clifford algebra for real Majorana basis.

i) For class D with no symmetry: The relevant Clifford algebra without reflection is:
\begin{equation}
\label{eq:2}
\{\gamma_i,M\}
\end{equation}
 The relevant Clifford algebra reads $\{\gamma_i,M,\gamma_x R_x\}$.

ii) For class with only TRS, the relevant Clifford algebra without reflection symmetry reads 
\begin{equation}
\label{eq:3}
\{\gamma_i, T, M\}
\end{equation}. If $[R_x,T]=0$, $\gamma_x R_x$ serves as a new generator. If $\{R_x,T\}=0$, $\gamma_x R_x T$ commutes with all original generators. This would not alter the Clifford algebra or change $Cl_{p,q}$ to $Cl_{p+q}$ depending on the square of the additional element.

iii) U(1)$\rtimes T$: Clifford algebra without reflection: 
\begin{equation}
\label{eq:4}
\{\gamma_i,T,TQ,M\}
\end{equation} We could add $\gamma_x R_x (R_{+}),\gamma_x R_x Q (R_-)$ to be another generator.

iv) U(1)$\rtimes [Z_2^C\times T]$ ($\{Q,T\}=0$): Clifford algebra without reflection: 
\begin{equation}
\label{eq:5}
\{\gamma_i,T,TQ,TQC,M\}
\end{equation} 
We could add $\gamma_x R_x (R_{++}),\gamma_x R_x Q(R_{--})$ to be another generator. Or the generator $\gamma_x R_x TC(R_{-+}),\gamma_x R_x TQC(R_{+-})$ commutes with all the original generators.

v) $SU(2)\times T$ ($[Q,T]=0$): Clifford algebra without reflection: 
\begin{equation}
\label{eq:6}
\{\gamma_i,TC,TQ,TQC,M\}
\end{equation}
 We could add $\gamma_x R_x (R_{++}),\gamma_x R_x Q(R_{--})$ to be another generator. Or the generator $\gamma_x R_x T(R_{-+}),\gamma_x R_x TQ(R_{+-})$ commutes with all the original generators. [Here we use the anti/commutation relation of $R_x$ with $\tilde T=TC$ to define the scenarios.]

vi) $SU(2)$ or U(1)$\rtimes Z_2^C$: Clifford algebra without reflection reads 
\begin{equation}
\label{eq:7}
\{\gamma_i Q,C,QC,QM\}
\end{equation} We could add $\gamma_x QR_x(R_+),\gamma_x R_x(R_-)$ to the original Clifford algebra.

vii) For the complex classes with U(1) generator $Q$, after choosing the basis where $Q$ reads $\sigma_2\otimes \mathds 1$, the kinetic and mass terms (time reversal $T$) are represented as a generator in the complex Clifford algebra\cite{lu2014}.
\begin{equation}
\label{eq:8}
\{\gamma_i,M,(T)\}
\end{equation}
we could add $\gamma_xR_x$ for A,AIII($R_+$) to the complex algebra or $\gamma_x R_xTQ$ for AIII($R_-$) commutes with the original generators.

The relevant Clifford algebra obtained as stated above is summarized at the second in the third column ``Clifford Algebra" in Table \ref{tab:results}.

For the case with two-fold rotation symmetry $U$ along the $x_d$ direction, we note that the elements defined by
\begin{equation}
S=U\prod _{i=1}^{d-1}\tilde\gamma_i
\end{equation} 
(anticommutes)commutes with all kinetic matrices $\tilde\gamma_i$'s and mass matrix $\tilde\beta$ in (even)odd spatial dimensions. Depending on its specific relation with global symmetries, the element $S(Q)(T)$ could either serve as another generator of the original Clifford algebra or commutes with all original generators as defined for Majorana basis above in \cref{eq:2,eq:3,eq:4,eq:5,eq:6,eq:7,eq:8}.

\section{The connection between $d$-dimensional reflection-SPT phases and $d-1$-dimensional SPT phases with internal $Z_2$ symmetry}
\label{appendixC}

\subsection{Strategy overview}

We work in the complex fermion basis below. A noninteracting topological phase in $d$-dimensional space is represented by the many-body ground state of the massive Dirac Hamiltonian (with respect to some particular choice of particle creation/annihilation fermionic operators)
\begin{equation}
\mathcal{H}=\int d^d \boldsymbol{x} \bf{\psi^\dagger (\bf x)}(\sum_i -i\partial_i \tilde\gamma_i+m\tilde \beta) \bf{\psi (\bf x)}
\end{equation}
consisting of mutually anticommuting hermitian matrices where the first terms represent the kinetic contribution and the second one is the mass term ($m\in \mathbb{R}$). In addition, the Hamiltonian may commute/anticommute with some anti-unitary operator which we denote as time-reversal ($\mathcal {T}$)/particle-hole symmetry ($\mathcal{C}$), respectively. There might exist an additional unitary symmetry that anticommutes with the Hamiltonian as chiral symmetry $\Gamma$. One could analyze the topological properties of this ground state by taking the stability analysis of the corresponding edge theory, i.e., the gapless edge modes on the interface between two phases which are generated by a ``domain wall'' configuration in the mass term, can be gapped if and only if these two phases could be connected without breaking any existing symmetries or closing the bulk gap. 

To construct the edge theory, we could write the Dirac mass term as $m_0sgn(z)M$ which is used to distinguish two topologically-inequivalent phases. The state $e^{-|m_0||z|}\chi$ where $\chi$ is the eigenvector that satisfies $i\tilde\beta\tilde\gamma_z \chi=sgn(m_0)\chi$ describes an edge mode that's localized to the domain wall in $z$ direction. The boundary Hamiltonian containing the dynamics of the edge modes is thus obtained by projecting the other kinetic terms (except $z$ direction) onto the subspace consisting of the eigenvectors with one certain eigenvalue of $i\tilde\beta\tilde\gamma_z $ which commutes with these kinetic $\tilde\gamma$ matrices, as well as time-reversal/particle-hole symmetry operators (if exist). We write the boundary Hamiltonian as 
\begin{eqnarray}
\mathcal H_{surface}=\sum_{i\neq z} -i\partial_i \gamma_i
\end{eqnarray}
 ($\gamma_i$'s denote the projected matrix of the original kinetic matrices from now on).

Now we assume the original Hamiltonian also possesses an additional reflection symmetry  in $x$ direction $R_x$ satisfying
\begin{eqnarray}
R_x^2=1, \{R_x,\tilde\gamma_x\}=0, [R_x,\tilde\gamma_i (i\neq x)]=0, 
 [R_x,\tilde\beta]=0.
\end{eqnarray}
The boundary Hamiltonian in the previous paragraph inherits all the symmetries and their corresponding algebraic relations from the original model. As conceived by Isobe and Fu, if we add another spatially-dependent mass term $m(x)e_m$ where $m(x)=m_0sgn(x)$ and $\{e_m,R_x\}=0$ that preserves all symmetries (the existence of the matrix $e_m$ will be discussed below in Sec \ref{odd_mass}), the low-energy degrees of freedom are confined to the domain wall where the gapless chiral edge modes lie. Therefore, if one manages to gap out the $d-2$ dimensional edge modes, the entire boundary of the original Hamiltonian is gapped. 
We write the boundary Hamiltonian with the reflection-odd mass $e_m$ as 
\begin{eqnarray}
\label{eqn:Hamiltonian}
\mathcal H_{sur,d-1}=\sum_{i\neq z} -i\partial_i \gamma_i+m(x)e_m
\end{eqnarray}

One can further obtain the $d-2$ dimensional boundary hamiltonian governing the chiral edge modes by similar procedure. Next, inspired by the idea of Isobe and Fu in Ref.\onlinecite{isobe2015}, we demonstrate that for certain cases, the $d-2$ dimensional edge theory could also be obtained as the edge theory of a $d-1$ dimensional system with all symmetries except that we substitute an internal symmetry for the spatial reflection symmetry (the algebraic relations, nevertheless, stay invariant). If the above statement holds, this will yield insight into the classification of reflection-symmetry protected topological phases using that of internal SPT phases in system with one dimension fewer.

\subsection{Equivalence of $d$-dimensional reflection SPT and $d-1$-dimensional $Z_2$ SPT phases}
We first choose a particular basis, where the operator we use to construct the edge modes $ie_m\gamma _x (\equiv E)$ is represented as $\mathds {1}\otimes \sigma_z$, namely we block diagonalize $E$ into its eigen subspace(choosing an orthonormal basis vectors that have eigenvalue $+1$ as $|1\ket,|2\ket,\cdots$). We further denote the basis as $|1\ket, |2\ket \cdots, \gamma_x |1\ket, \gamma_x |2\ket \cdots$ since $\{\gamma_x, E\}=0$. So we also fix $\gamma_x$ as $\mathds{1} \otimes \sigma_x$ and $e_m=-iE\gamma_x=\mathds{1}\otimes \sigma_y$. All kinetic matrices other than $\gamma _x$ as well as other symmetries $\mathcal T,\mathcal C,\Gamma$ are block diagonalized in this basis since they commute with $E$. Since $\bra n|\gamma_x \gamma_i \gamma_x|n'\ket=-\bra n|\gamma_i|n'\ket$, the other kinetic matrices can be represented as $\Gamma_i\otimes \sigma_z$. Similarly, $\mathcal T,\mathcal C,\Gamma,R_x$ is represented as $e_T\otimes \sigma_z \mathcal K, e_C\otimes \sigma_0 \mathcal K, e_\Gamma\otimes\sigma_z, e_R\otimes \sigma_zP$ (here $P$ denotes the operation in real space that changes $x$ to $-x$. $e_T,e_C,e_\Gamma,e_R$'s simply denote some Hermitian matrix acting on the remaining degrees of freedom). Under this choice of basis, the $d-1$ dimensional surface Hamiltonian \ref{eqn:Hamiltonian} reads 
\begin{eqnarray}
\label{eqn:Hamiltonian2}
\mathcal H_{sur,\!d\!-\!1}\!=\!-i\partial_x \mathds{1}\!\otimes\! \sigma_x+\!\sum_{i\neq x,z}\!-i\partial_i \Gamma_i\!\otimes\! \sigma_z+m(x)\!\mathds{1}\!\otimes\!\sigma_y\nonumber\\
\mathcal T=e_T\otimes \sigma_z\mathcal K\text{(if exists)}, \mathcal C=e_C\otimes \sigma_0\mathcal K\text{(if exists)}\nonumber\\ \Gamma=e_\Gamma\otimes\sigma_z\text{(if exists)}, R_x=e_R\otimes\sigma_z P
\end{eqnarray}
and the $d-2$ dimensional boundary Hamiltonian can be expressed as 
\begin{eqnarray}
&&\mathcal H_{bd,d-2}=\sum_{i\neq x,z} -i\partial_i \Gamma_i \nonumber\\
\mathcal T&&=e_T\mathcal K \text{(if exists)}, \mathcal C=e_C\mathcal K \text{(if exists)}, R_x=e_R
\end{eqnarray} 
(Note that $R_x$ no longer contains real space operator $P$ and is an on-site symmetry in the edge theory). 

If we interpret Hamiltonian \eqref{eqn:Hamiltonian2} as describing a $d-1$ dimensional system with the same time-reversal and/or particle-hole symmetries, albeit the reflection operation $R_x=e_R\otimes \sigma_zP$ is changed to a new operator $g=e_R\otimes \sigma_0$. This alteration, notwithstanding, won't revise the algebraic relation of $\mathcal T,\mathcal C, \Gamma, \gamma_i (i\neq x)$ with $g/R_x$, yet it will make $\gamma_x, e_m$ commute with $g$. So now $g$ serves as an internal symmetry operator that shares the same algebraic relation with other symmetries as $R_x$. 
\begin{equation}
\begin{split}
\label{eqn:Hamiltonian2}
\mathcal H_{sur,\!d\!-\!1}\!=\!-i\partial_x \mathds{1}\!\otimes\! \sigma_x+\!\sum_{i\neq x,z}\!-i\partial_i \Gamma_i\!\otimes\! \sigma_z+m\!\mathds{1}\!\otimes\!\sigma_y\nonumber\\
\mathcal T=e_T\otimes \sigma_z\mathcal K\text{(if exists)}, \mathcal C=e_C\otimes \sigma_0\mathcal K\text{(if exists)}\\ \Gamma=e_\Gamma\otimes\sigma_z\text{(if exists)}, g=e_R\otimes\sigma_0
\end{split}
\end{equation}
The edge theory obtained from this $d-1$ dimensional system with the same domain wall configuration in the mass term is the same as that of the $d$ dimensional system. The interaction terms that gap out this $d-1$ dimensional system boundaries therefore also respect all the symmetries of the $d$-dimensional system.

This connection would yield an upper bound for the $\mathbb Z_n$ classification of $d$ dimensional SPT phase. Next we show that this is the case for the example illustrated in the paper by Isobe and Fu (which is later elaborated on by Yoshida and Furusaki). Written in the basis $|+y\ket_0,|-y\ket_0,(\gamma_x=)\sigma_y\otimes \sigma_0|+y\ket_0,\sigma_y\otimes\sigma_0|-y\ket_0$ as demonstrated in eqn (31b), eqn(31c) in Ref.\onlinecite{yoshida2015}, the surface Hamiltonian of the 3d TCI  eqn(30) can be expressed as 
\[ \mathcal H_{sur,d-1}=(i\partial_x \sigma_x-i\partial _y \sigma_z)\otimes\sigma_z+m(x)\sigma_0\otimes\sigma_y\]
and the symmetries are $\mathcal T=-i\sigma_y\otimes \sigma_z\mathcal K, R_x=i\sigma_z\otimes\sigma_z$. Similarly, for the 2d system Hamiltonian eqn(1), written in the basis [eqn (4b)]$|\pm y\ket_0,(\gamma_x=)\sigma_y\otimes\sigma_z|\pm y\ket_0$, it reads the same as the above Hamiltonian with the symmetries $\mathcal T=i\sigma_y\otimes\sigma_z\mathcal K, g=i\sigma_z\otimes \sigma_0$. So the only difference is between symmetries $g,R_x$ where we change $\sigma_z$ in $R_x$ to $\sigma_0$ in $g$. Thus the classification of the 3D TCI $\mathbb Z_8$ is given by the $\mathbb Z_4$ classification of the 2D model with internal symmetry $g$. (The difference of factor two originates from the fact that in order to find the $e_m$ matrix in the surface Hamiltonian of the 3D system, we have to enlarge the dimension of the matrix by two which means using two copies of the surface.)
\subsection{ Existence of $e_m$}
\label{odd_mass}

Next we will discuss exhaustively whether the mass term $e_m$ in Eq.\eqref{eqn:Hamiltonian} exists for each symmetry class and different commutation relations with reflection symmetry. Define $R_x\mathcal T=\eta_T \mathcal T R_x$ and $R_x\mathcal C=\eta_C \mathcal C R_x$ where $\eta_T,\eta_C$ are $\pm 1$. We first note that in the $d-1$ surface Hamiltonian of the original $d$-dimensional system, the term $i\gamma_x R_x$ already anticommutes with other kinetic gamma matrices as well as $R_x$ itself (i.e., already satisfy the algebraic relation of $e_m$ with these terms), we only need to make it consistent with other protecting symmetries of the symmetry class. If it falls into one of the following three scenarios

(i)the unitary class A

 (ii)there's only one anti-unitary symmetry ($\mathcal T/\mathcal C$) 
 
 (iii)there're two anti-unitary symmetries and the algebraic relations of reflection with the two anti-unitary symmetries (for real chiral symmetry class) are \emph {the same} (i.e., $\eta_T=\eta_C$), 
 
then we can \emph {always} manage to render the above term consistent with other protecting symmetry(ies) by leaving it intact [for the case where reflection commutes with the symmetry(ies)] or tensor producting it with $\sigma_y$ to reverse its original (anti-)commutation relation with protecting symmetries [in the case where reflection anti-commutes with protecting symmetry(ies), note that $\mathcal T/\mathcal R$ are anti-unitary]. We could confirm that this is indeed the case in Ref. \onlinecite{yoshida2015} where the original representation for their three dimensional surface Hamiltonian contains [eqns (27),(29) of Ref.\onlinecite{yoshida2015}]\[R_x\sim \sigma_x P, \mathcal T\sim \sigma_y \mathcal K, \gamma_x=\sigma_y\] with $\{R_x,\mathcal T\}=0$ and that the additional mass term \[e_m=\sigma_z\otimes\sigma_y\sim i R_x \gamma_x\otimes \sigma_y.\]

The above discussion leaves out two scenarios: 

(i) the chiral complex class AIII; 

(ii) the chiral real class with $\eta_T\eta_C=-1$. 

We first prove that the above SPT equivalence doesn't apply to the case  for chiral complex class when the reflection symmetry anti-commutes with chiral symmetry $\Gamma$ and the case $(\eta_T,\eta_C)=(1,-1)$ for class BDI and CII as well as $(\eta_T,\eta_C)=(-1,1)$ for class DIII and CI [there're two possibilities accounting for the ineffectiveness, either because of the \emph{non-existence} of $e_m$ or the original reflection-protected classification is already trivial/$\mathbb Z_2$ yet we need to enlarge the dimension by two to construct such a matrix which means that this equivalence relation won't modify the original classification scheme].  

We relax the restriction that $e_m$ must anti-commute with $R_x$ first [We could infer about this by examining the noninteracting classification: If the noninteracting classification is $\mathbb Z$ and yet we find such a mass term then it's guaranteed that it anti-commutes with reflection. If the original noninteracting classification is already trivial/$\mathbb Z_2$, this equivalence relation won't give information about the collapse of the classification.]. According to Appendix \ref{clifford}, the addition of reflection symmetry on the original Hamiltonian doesn't alter the associated clifford algebra, so the classification is actually the same as the original ``AZ" classes without reflection symmetry. The presence of $e_m$ corresponds to the gapping of the surface Hamiltonian. If the classification is $\mathbb Z$, then no such mass term exists in the surface Hamiltonian irrespective of its relation with reflection symmetry; if it's $\mathbb Z_2$, we have to use two copies of the system to gap out the surface Hamiltonian, etc. So in these cases the equivalence relation we find won't yield meaningful outcome for the collapse of the classification. 

For the remaining possibilities 

(i)AIII when reflection commutes with $\Gamma$, 

(ii)$(\eta_T,\eta_C)=(-1,1)$ for class BDI and CII,

(iii) $(\eta_T,\eta_C)=(1,-1)$ for class DIII and CI.

we could determine the existence of the reflection-odd mass term $e_m$ in the surface Hamiltonian as following: first we determine the rank of root state for a certain scenario using Clifford algebra. Their relevant clifford algebras in the presence of the reflection symmetry are $Cl_{d+3}$ for AIII with commuting reflection symmetry and $Cl_{d+4}$ for the last four real chiral symmetry classes. We denote the dimension of its surface Hamiltonian (which is half of that of the bulk) as $r_{sur}$. Then we denote the dimension of the root state of the Hamiltonian in the same symmetry class albeit \emph {without} reflection symmetry in $d-1$-dimensional system as $r_1$. The complete clifford algebras for $d-1$ dimensional systems without reflection symmetry are $Cl_{d+1}$ for AIII and $Cl_{p,q}=Cl_{d,3}/Cl_{d+2,1}/Cl_{d-1,4}/Cl_{d+1,2}$  for symmetry classes BDI/CII/DIII/CI, respectively. If $r_{1}\leq r_{sur}$, then we are sure to find such a mass term, which is the case for AIII by virtue of $Cl_{n+2}\cong Cl_n\otimes_{\mathbb C} \mathbb C(2)$; otherwise, if $r_{sur}=\frac{r_1}{2}$, we switch to find the minimal dimension upon trading the mass term in the Hamiltonian for a kinetic term [namely the rank of $Cl_{p+1,q-1}$] and denote it as $r_2$. If $r_2=r_{sur}$, this means we can find an additional kinetic term in the representation of the surface Hamiltonian, then by tensoring this with $\sigma_y$ we can make it a legitimate mass term. If all the above procedure fails to yield a mass term, then it's impossible to find one. By this algorithm with some calculation, we conclude that the mass term \emph{doesn't} exist for $d=8n+5/8n+1/8n+7/8n+3$ dimension systems for symmetry classes BDI/CII/DIII/CI, respectively, with the abovementioned reflection symmetries. While in other dimensions for real chiral classes as well as for AIII in all dimensions with commuting $R_x$, a mass term is sure to exist and we could exploit this equivalence to extract information of the collapse. 

\bibliographystyle{apsrev4-1}
\bibliography{breakdown_v15}
\end{document}